\documentclass[11pt,a4paper]{article}
\pdfoutput=1
\usepackage{jheppub}
\usepackage{latexsym,amsfonts,amsmath,amssymb}
\usepackage{graphicx}
\usepackage{color}
\usepackage{caption}
\usepackage{subcaption}
\usepackage{url}
\usepackage{slashed}
\newcommand{\CN}{\mathcal{N}}

\newcommand{\ndt}{\noindent}

\newcommand{\nn}{\nonumber}

\def\p{\partial}

\def\bea{\begin{eqnarray}}
\def\eea{\end{eqnarray}}
\def\be{\begin{equation}}
\def\ee{\end{equation}}
\def\bse{\begin{subequations}}
\def\ese{\end{subequations}}

\newcommand{\bem}{\begin{pmatrix}}
\newcommand{\eem}{\end{pmatrix}}

\renewcommand{\=}{\;  = \;}
\def\+{\, + \,}

\def\wt{\widetilde}

\def\bar{\overline}

\def\rt2{\sqrt{2}}

\def\s{\sigma}
\def\g{\gamma}

\def\a{\alpha}

\def\eps{\epsilon}

\def\m{\mu}
\def\n{\nu}

\def\nv{n_\text{v}}

\title{On the localization manifold of 5d supersymmetric spinning black holes}

\author{Rajesh Kumar Gupta,}
\author{Sameer Murthy,}
\author{Manya Sahni}

\affiliation{Department of Mathematics, King's College London,\\
  The Strand, London WC2R 2LS, U.K}

\emailAdd{rajesh.gupta@kcl.ac.uk}
\emailAdd{sameer.murthy@kcl.ac.uk}
\emailAdd{manya.sahni@kcl.ac.uk}

\abstract{
We analyze the localization equations relevant to the quantum entropy of spinning supersymmetric 
black holes in five-dimensional asymptotically flat space. 
The precise problem is to classify all solutions to the off-shell supersymmetry equations in~$\CN=2$ supergravity 
coupled to~$\nv+1$ vector multiplets around the near-horizon black hole. 
We rewrite these equations in terms of the bosonic spinor bilinears that exist in the geometry for an arbitrary background. 
We then focus on the vector multiplet fluctuations around the near-horizon attractor region of 
the supersymmetric black hole, and classify all smooth solutions to the localization equations in this background
for different choices of analytic continuation. 
For the choice of analytic continuation consistent with the 4d/5d lift, 
we find that the most general localization solution for the five-dimensional black hole problem is an~$(\nv+1)$-dimensional
manifold, which is precisely the lift of the localization manifold for supersymmetric black holes in 
four-dimensional asymptotically flat space.
}

\begin{document}

\maketitle

\section{Introduction \label{sec:introduction}}

There has been good progress in the last twenty years in formulating and computing the 
quantum entropy of supersymmetric black holes. This is a notion that extends the semiclassical 
Bekenstein-Hawking entropy, which is valid in the thermodynamic limit when the black hole size is infinite, 
to a quantity that is defined when the curvatures and coupling constants cannot be ignored. 
Starting from the work of~\cite{LopesCardoso:1998tkj}, developments in this direction 
have led to the formulation of the quantum black hole entropy as a functional 
integral over all the fields of the gravitational theory in the Euclidean near-horizon~AdS$_2$ region of the 
black hole~\cite{Sen:2008vm}. In this formulation, the corrections to the Bekenstein-Hawking formula
are split into two types of effects. The first type consists of corrections that can be encoded in the  
effective action as the inclusion of local higher-dimension operators---these can be recovered as a saddle point 
approximation to the functional integral. The second type consists of quantum corrections that  
arise from loop effects around the~AdS$_2$ background. 

The leading logarithmic one-loop quantum corrections to the semiclassical entropy has been computed in a variety of 
situations (see~\cite{Mandal:2010cj} for a summary). In situations with supersymmetry one can go further and try to compute the 
full functional integral, using the technique of localization, to obtain the \emph{exact} quantum 
entropy~\cite{Dabholkar:2010uh,Dabholkar:2013,Dabholkar:2014ema}. 
There are, of course, many subtleties and difficulties in applying localization to supergravity, but the basic ideas seem 
to be in good enough shape by now to be applied to various situations. In particular, we now understand how to formulate 
a notion of a rigid off-shell~$Q$ which acts covariantly on all the fields of the theory, using an adaptation of the background field 
method~\cite{deWit:2018dix, Jeon:2018kec}.
One starts with a supersymmetric background to an off-shell supergravity theory---the attractor black hole configuration 
in our situation---with one choice of 
Killing spinor, and the problem then reduces to finding all configurations of the supergravity theory which admit 
\emph{some} Killing spinor that asymptote to the attractor background in a manner that the Killing spinor 
asymptotes to the chosen background Killing spinor. The off-shell nature of the supersymmetry variations 
allows us to consider the problem separately in the Weyl and matter multiplets of the theory. One lists the set of all 
supersymmetric matter fluctuations around each supersymmetric Weyl multiplet configuration, and  
the localization manifold consists of the combined space of solutions.

Most of the progress in applying these ideas to exact quantum black hole entropy
has been in the context of supersymmetric black holes in~$\CN=2$ supergravity 
theories in four-dimensional asymptotically flat 
space~\cite{Gupta:2012cy,Murthy:2013xpa,Gupta:2015gga,Murthy:2015yfa,Murthy:2015zzy}, 
leading up to a gravitational derivation of the OSV conjecture~\cite{Ooguri:2004zv} in its sharpened form~\cite{Denef:2007vg}. 
It is natural to ask if this progress can be extended to supersymmetric black holes in higher dimensions. 
In this paper we study this problem for a class of spinning black holes in five-dimensional asymptotically flat space, 
namely the BMPV black holes~\cite{Breckenridge:1996is}. These 5d black holes are intimately related to the 4d black holes
mentioned above via the 4d/5d connection~\cite{Gaiotto:2005gf}. 
Indeed, embedding the BMPV black holes in M-theory compactified
on a Calabi-Yau 3-fold and placing this configuration on the tip of a Taub-NUT space brings us to a 4d black hole solution 
in Type IIA string theory on the same Calabi-Yau.
The spin of the 5d black hole become angular momentum around the Taub-NUT space
which is seen as a gauge charge by the 4d black hole.

In the context of the current paper we regard the 5d supersymmetric spinning BMPV black holes as solutions to 
5d off-shell~$\CN=2$ supergravity coupled to~$\nv+1$ vector 
multiplets~\cite{Bergshoeff:2001hc,Fujita:2001kv, Bergshoeff:2004kh,Hanaki:2006pj}.
In the near-horizon attractor region, the metric is 
fixed to~$AdS_2 \times S^2 \ltimes S^1$, where the~$S^1$ is fibered over the~$AdS_2$ 
as well as the~$S^2$, and the vector fields are fixed to have constant electric field strengths~\cite{deWit:2009de}.
In this paper we perform a complete analysis of the localization manifold in the vector multiplet sector 
with attractor boundary conditions. We also begin an analysis of the general solution in the Weyl multiplet sector, 
but we postpone an exhaustive analysis to future work.\footnote{We follow 
this strategy partly because the analysis in the Weyl multiplet sector is 
technically quite intricate, and partly because a similar strategy in 4d led to rapid progress in understanding the 
quantum entropy~\cite{Dabholkar:2010uh}. In the 4d situation, it was then shown rigorously that essentially the only BPS solution 
in the Weyl multiplet sector with the attractor boundary conditions is the (fully supersymmetric) attractor solution 
itself~\cite{Gupta:2012cy}.}

The problem of finding the localization manifold in the vector multiplet sector 
has been addressed previously in~\cite{Gomes:2013cca} 
making heavy use of the \emph{off-shell 4d/5d connection}~\cite{Banerjee:2011ts}.
The off-shell 4d/5d connection spells out a precise relation between solutions of off-shell BPS equations in 5d with their 
4d counterparts. The basic statement is that of a consistent truncation of the off-shell BPS equations: 
off-shell BPS configurations in 4d lift to off-shell BPS configurations in 5d. 
In the black hole context, the complete set of solutions in 4d has been solved in~\cite{Dabholkar:2010uh,Gupta:2012cy}, 
and the solution set consists of a one-parameter family in each vector multiplet, thus yielding a (real) $(\nv+1)$-dimensional manifold.
Integrating the action of this manifold 
then leads to an OSV-type integral formula for the quantum entropy. Putting together the 4d/5d black hole connection~\cite{Gaiotto:2005gf} 
with the off-shell 4d/5d connection, the work of~\cite{Gomes:2013cca} showed that the above 4d solutions can be lifted to 5d solutions 
around the black hole attractor.  
More precisely, the analysis of~\cite{Gomes:2013cca} showed that if we switched off some of the auxiliary fields in the 5d vector multiplet,
the remaining fluctuations are constrained by a set of equations similar to the contact instanton equations arising in~\cite{Kallen:2012va}.

The obvious question that arises is whether there are any other new solutions to the BPS equations. 
Firstly, there could be non-trivial smooth solutions to the contact-instanton-like equations. 
Secondly, there may be more solutions to the localization equations (potentially an infinite number) if we switch on all the auxiliary fields.  
More solutions to the localization equations would imply that the localization approach to quantum entropy of 5d black holes 
is much more complicated than the corresponding 4d story. While this conclusion is a technical possibility---after all, the equations 
of 5d localization are more complicated because they involve a fifth direction in which all the fields fluctuate---it seems a bit at odds with 
the fact that the near-horizon configurations of the 4d and 5d black holes are very closely related by the 4d/5d lift. 
It also seems to be at odds with the fact that the microscopic ensembles, as well as the expressions for the microscopic degeneracies, 
are almost the same~\cite{Dijkgraaf:1996xw,Castro:2008ys,Dabholkar:2010rm}.

These considerations led us to revisit this problem of finding the localization manifold in the 
vector multiplet sector for 5d black holes including all the 5d fluctuations of all the auxiliary fields in the theory. 
The approach we take is to directly analyze the 5d off-shell BPS equations 
instead of going through the 4d/5d lift. 
We find that the solution manifold depends strongly on the Euclidean continuation that is used. 
We explore different choices that could each be termed natural from certain points of view.
We find that one choice (that we call~A1) leads to a finite-dimensional manifold which is precisely the lift of the 
4d localization manifold found in~\cite{Dabholkar:2010uh,Gupta:2012cy} with no extra 
solutions.\footnote{We emphasize that smoothness of the solutions is an essential criterion in 
reaching this conclusion. This leaves open the possibility of instanton-like (or orbifold) solutions, which
are known to play an important role~\cite{Banerjee:2008ky,Murthy:2009dq, Dabholkar:2014ema}.}
This choice is precisely the one that reduces to the 
analytic continuation of the 4d theory used in~\cite{Dabholkar:2010uh,Gupta:2012cy}. 
We also present the results for other choices of analytic continuation (called A2, B, C),
including one which preserves~5d covariance, partly because these solutions could be relevant 
for some other physical problem. 

While our main results concern the vector multiplet localization locus for the black hole quantum entropy problem, 
we also begin a treatment of the Weyl mutliplet for arbitrary backgrounds. 
We use the method of fermion bilinears~\cite{Tod:1983pm, Gauntlett:2001ur, Gauntlett:2002sc} 
to reduce the off-shell problem to a system of coupled PDEs among bosonic 
quantities, and find the most general (local) solution. 
We do this both in the Weyl and vector multiplets. The next step, which is  
to impose the attractor boundary conditions and perform a careful Euclidean continuation\footnote{A Euclidean 
continuation is required because our eventual goal is to compute functional integrals. Ideally we should 
begin with a Euclidean formulation of our theory, but unfortunately we do not know of a suitable Euclidean 
supergravity formulation. In the 4d case a similar route was followed at first in~\cite{Dabholkar:2010uh,Dabholkar:2013}, 
and a Euclidean formalism was later developed in~\cite{deWit:2017}, which confirmed the solutions found by Euclidean continuation.}, 
can be considerably difficult in the Weyl multiplet because of the mixing of many fields of varying spins (see e.g.~\cite{Gupta:2012cy}). 
We postpone the full analysis of this interesting problem to future work.

The plan of the paper is as follows.
In Section~\ref{sec:sugraBH} we review the formalism of 5d $\CN =2$ off-shell supergravity, the supersymmetric 
spinning black hole solutions, and the off-shell 4d/5d connection. 
In Section~\ref{sec:Weyl} we review the method of fermion bilinears and present the most general local
solution in the Weyl multiplet sector. 
In Section~\ref{sec:Vector} we find the most general local solution in the vector multiplet sector. We then
apply boundary conditions and discuss different Euclidean continuations and find the most general 
solution with attractor boundary conditions. 
In Section~\ref{sec:Discussion} we discuss how our results fit into the quantum entropy program in string theory
as well as the future directions to be attacked. 
In two appendices we summarize our conventions and present the Killing spinors used in the main text.

\section{5d $\CN=2$ supergravity and spinning black holes \label{sec:sugraBH}}

In this section we briefly review some relevant aspects of~$\CN =2$ off-shell supergravity coupled to 
vector multiplets in 5 dimensions, following the construction based on the superconformal 
formalism~\cite{Bergshoeff:2001hc,Fujita:2001kv, Bergshoeff:2004kh,Hanaki:2006pj,deWit:2009de}. 
We then review the spinning supersymmetric black hole solutions in this theory. Finally we briefly review 
the off-shell 4d/5d connection of~\cite{Banerjee:2011ts}.

\subsection{Off-shell 5d $\CN=2$ supergravity coupled to vector multiplets \label{sec:offshell}}

In five space time dimensions the independent fields in the Weyl multiplet are 
\be\label{WeylFields}
\mathbf{W} \= (e_{M}^{A}, \psi_{M}^{i}, b_{M}, V_{M i}{}^{j}; T_{AB}, \chi^{i}, D) \,.
\ee
Here the indices $\{A,B\}$, $\{M,N\}$, and $\{i,j\}$ are five dimensional flat space, curved space, and $SU(2)_{R}$ indices, respectively. 
The fields are the f${\ddot{\text u}}$nfbein $e_{M}^{A}$, the gravitini~$\psi_{M}^{i}$, dilatation 
gauge field $b_{M}$, $SU(2)_{R}$ gauge fields $V_{M i}{}^{j}$, as well as the auxiliary 
fields which are the anti-symmetric tensor~$T_{AB}$, the spinor~$\chi^{i}$, and the scalar $D$.    
In addition the Weyl multiplet also includes the fields $\omega_{M}{}^{AB}, f_{M}{}^{A},\phi_{M}{}^{i}$,
which are gauge fields corresponding to Lorentz transformations, special conformal transformations, 
and special supersymmetry transformations, respectively. These fields are determined in terms of 
the independent fields of the multiplet \eqref{WeylFields} through conventional constraints, and are termed 
composite.
The supersymmetry variation of the gravitini is given by
\be \label{WeylVar5d}
\delta\psi^{i}_{M} \=2\mathcal D_{M}\epsilon^{i}+\frac{i}{2}T_{AB}(3\g^{AB}\g_{M}-\g_{M}\g^{AB})\epsilon^{i}-i\g_{M}\eta^{i}\,,
\ee
where~$\epsilon^{i}$ and~$\eta^{i}$ parameterize the supersymmetry ($Q$) and conformal supersymmetry ($S$) respectively.
The operator~$\mathcal D_{M}$ is the covariant derivative with respect to all superconformal transformations except the 
special conformal transformations, and it acts on the supersymmetry parameter as 
\be
\mathcal D_{M}\eps^{i} \= \bigl(\p_{M}-\frac{1}{4}\omega_{M}{}^{AB}\g_{AB}+\frac{1}{2}b_{M} \bigr) \, 
\eps^{i}+\frac{1}{2}V_{M j}{}^{i}\eps^{j} \,.
\ee

The five-dimensional vector multiplet is 
\be
\mathbf{V} \= (\sigma, \Omega^{i}, W_{M}, Y^{ij}) \,.
\ee
The components of this multiplet are the gauge field $W_{M}$, the real scalar $\sigma$, the gaugini $\Omega^{i}$, which is a doublet under $SU(2)_{R}$,  
and auxiliary field $Y^{ij}$ which is a triplet under~$SU(2)_R$.  
The supersymmetry variation of the gaugini is
\be \label{VectorSusy}
\delta\Omega^{i} 
\=-\frac{1}{2}(\widehat F_{AB}-4\sigma T_{AB})\g^{AB}\eps^{i}-i\slashed D\sigma\eps^{i}-2\varepsilon_{jk}Y^{ij}\eps^{k}+\sigma\eta^{i}\,.
\ee

The anticommutator of supersymmetry transformations close on to the algebra generated by the 
general coordinate transformations, Lorentz transformation and gauge transformations.

The action of the Weyl multiplet coupled to~$\nv$ vector multiplets is 
\be\label{5dLagr.}
\begin{split}
8\pi^{2}\mathcal L&\=3C_{IJK}\sigma^{I}\Big[\frac{1}{2}\mathcal D_{\m}\sigma^{J}\mathcal D^{\m}\sigma^{K}+\frac{1}{4}F_{\m\n}{}^{J}F^{\m\n K}-Y_{ij}{}^{J}Y^{ij K}-3\sigma^{J}F_{\m\n}{}^{K}T^{\m\n}\Big]\\
&-\frac{i}{8}C_{IJK}e^{-1}\varepsilon^{\m\n\rho\sigma\tau}W^{I}_{\m}F_{\n\rho}{}^{J}F_{\sigma\tau}{}^{K}-C(\sigma)\Big[\frac{1}{8}R-4D-\frac{39}{2}T^{2}\Big] \,.
\end{split}
\ee
Here $C_{IJK}$ is the symmetric tensor of~$\CN=2$ supergravity, and the function $C(\sigma)$ are defined as
\be
C(\sigma)\=C_{IJK}\sigma^{I}\sigma^{J}\sigma^{K}.
\ee

\subsection{Supersymmetric black holes in 5d \label{sec:susyBH}}
The action \eqref{5dLagr.} admits charged, rotating black hole solutions. This action is essentially an Einstein-Maxwell theory,
with an additional Chern Simons interaction for the gauge fields~$W\wedge F\wedge F$, which has the same dimensions 
as the Maxwell terms. The presence of the Chern Simons term does not affect static solutions, 
but it does affect stationary solutions. 

We consider asymptotically flat black holes characterized by mass $M$, charge $Q$ and two angular momenta $J_{1}$ and $J_{2}$. 
The existence of the horizon covering the singularity requires that the mass and charge of the black hole satisfy 
the following inequality, for any value of the angular momenta \cite{Gibbons:1993xt}
\be\label{Mass.ineqality}
M \, \geq \, \frac{\sqrt{3}}{2}|Q| \,.
\ee
When this inequality is saturated one linear combination~$J_R$ of~$J_{1}$ and~$J_{2}$ vanishes, 
and the black hole solution admits a Killing spinor. 
Thus, a 5d supersymmetric black hole is characterized by the charge~$Q$ and one angular 
momentu~ $J_L$~\cite{Breckenridge:1996is,Gauntlett:1998fz}. 

Supersymmetry implies that the near horizon geometry of such a supersymmetric black hole (charged and rotating) 
is completely fixed by the charges of the black hole, which is the statement of the attractor mechanism.
This geometry consists of the product of AdS$_{2}$ and S$^{2}$ of equal radii, and a circle which is 
non trivially fibered over AdS$_{2} \times$ S$^{2}$. 
This metric can be written as follows~\cite{deWit:2009de},
\be\label{NearHorzMetric}
ds^{2}\=\frac{1}{16v^{2}}\Big(-r^{4}dt^{2}+4\frac{dr^{2}}{r^{2}}+d\theta^{2}+\sin^{2}\theta\,d\phi^{2}\Big)+e^{2g}(d\psi+B)^{2}\,,
\ee
where $\psi$ is the periodic coordinate of the circle and 
\be
B\=-\frac{1}{4v^{2}}e^{-g}(T_{23}r^{2}\,dt-T_{01}\cos\theta\,d\phi)\,.
\ee
Here, $g$ and the tensor field~$T_{AB}$ are constants 
which determines the size of the circle and its fibration over the base space, respectively. The value of $T_{23}$ determines the angular 
momentum of the black hole. 
The parameter~$v$ is a constant that determines the size of~AdS$_{2} \times$ S$^{2}$. Using the dilatation symmetry of~$\CN=2$
supergravity we set $v=\frac{1}{4}$. 
The conditions for supersymmetry then constraint $T_{ab}$ to obey  
\be
(T_{01})^{2}+(T_{23})^{2}\= v^2 \=\frac{1}{16}.
\ee
For later convenience we will parametrize the background value of $T_{AB}$ as
\be \label{NearHorzT}
T_{01}\=\frac{1}{4}\cos\beta\,,\quad T_{23}\=\frac{1}{4}\sin\beta \,.
\ee
In the case $T_{23}=0$, the metric \eqref{NearHorzMetric} reduces to that of the charged static black hole with near-horizon 
geometry AdS$_{2}\times$S$^{3}$, whereas in the case of $T_{01}=0$ the metric reduces to that of the black string
with near-horizon geometry AdS$_{3}\times$S$^{2}$. 
Arbitrary values of~$T_{ab}$ correspond to spinning black holes with near-horizon geometry AdS$_{2}\times$S$^{2} \ltimes S^1$, 
and interpolates between these two limits.

The entropy of this black hole is given by the area of the 3-dimensional compact space at~$r=0$~\cite{deWit:2009de},  
\be
S_\text{BH} \=4\pi e^{g}.
\ee
For the vector multiplet fields the near horizon supersymmetry requires that the value of the scalar field is constant~i.e.,
\be\label{AttreactorSigma}
\sigma^{I}=\sigma^{I}_{*}\,, \qquad I\=1,..,n_{v} \,, 
\ee
with the constants~$\sigma^{I}_{*}$ are determined in terms of the charges using the fact that 
the gauge field strength is given in terms of $T_{ab}$ as 
\be\label{NearHorzGaugeFld}
F^{I}_{tr}\=4\sigma_{*}^{I}\,T_{01} \,,\qquad F^{I}_{\theta\phi}\=4\sigma_{*}^{I}\,T_{23} \sin\theta \,.
\ee
The near horizon field configurations given by the metric~\eqref{NearHorzMetric} and scalar fields~\eqref{NearHorzGaugeFld} 
preserves 8 real supercharges. The explicit form of the Killing spinors in the Euclidean version of this theory is given in Appendix~\ref{App:KS}.

\subsection{Off-shell 4d/5d connection}\label{4d5dOffShell}

It is well known that upon dimensional reduction from 5d to 4d, the equations of motions of the 5d theory 
reduce to the equations of motion of the 4d theory.
A slightly less known fact is that a similar statement is also true for the \emph{off-shell} dimensional reduction: 
off-shell BPS equations in 5d reduce to off-shell BPS equations in 4d~\cite{Banerjee:2011ts}. Therefore BPS configurations in 4d can be 
lifted to BPS configurations in 5d, and conversely, 5d BPS configurations which are independent of the fifth direction
reduce to 4d BPS configurations.

In our context, the conformal 5d supergravity coupled to $\nv$ vector multiplets that was discussed above reduces to the off-shell
conformal 4d supergravity coupled to $\nv+1$ number of vector multiplets discussed in~\cite{deWit:1984rvr}.
Upon dimension reduction the 5d Weyl multiplet reduces to the Weyl multiplet and a Kaluza Klein vector multiplet in 4d, 
a 5d vector multiplet reduces to a 4d vector multiplet. 
In particular, taking the standard Kaluza-Klein ansatz for dimensional reduction along a circle, the 5d and 4d vielbeins are related as 
\be
e_{M}{}^{A}\=\begin{pmatrix}e_{\m}{}^{a}\,\, &B_{\m}\,\phi^{-1}\\0\,\,&\phi^{-1}\end{pmatrix},\qquad 
e_{A}{}^{M}\=\begin{pmatrix}e_{a}{}^{\m}\,\,&-e_{a}{}^{\n}B_{\n}\\0\,\,&\phi\end{pmatrix} \,.
\ee
Here $\{a,b, \cdots\}$ and $\{\m,\n, \cdots \}$ are 4 dimensional flat-space and curved-space indices, respectively. 
In the above ansatz, $B_{\m}$ represents a non trivial fibration of the circle with the size~$\phi^{-1}$ over the 4d base space. 
The auxiliary~$T$ field reduces as follows, 
\be
T_{AB}\=\begin{pmatrix}T_{ab}\\-\frac{1}{6}A_{a}\end{pmatrix}\,.
\ee

In the dimension reduction of a vector multiplet, the 5d vector field reduces to a 4d vector field and a real scalar. 
The extra scalar combines with the scalar of the 5d vector multiplet to yield the complex scalar of the 4d vector multiplet. 
More explicitly, the 4d vector multiplet fields, $(X,\lambda_{i},A_{\m},y^{ij})$, in terms of 5d vector multiplet fields 
are given by, with~$W\equiv W_{5}$, 
\be \label{4d5dvec}
X^I \=-\frac{1}{2}i(\sigma^I+i\phi W^I)e^{-i\varphi}, \qquad A^I_{\m}\=W^I_{\m} \,, \qquad I = 1,\dots, \nv \,.
\ee 
The $I=0$ vector multiplet is built out of fields of the Weyl multiplet as follows,
\be
X^{0}\=-\frac{1}{2}\phi\,e^{-i\varphi}\, , \qquad A^{0}_{\m}\=B_{\m} \,.
\ee
Here $\varphi$ is a scalar field which transforms inhomogeneously under 4 dimensions chiral $U(1)$ gauge transformation 
for which the gauge field is $\mathcal A_{\m}$. This field is introduced so that the 4d fields have right chiral charges.
Comparing these results with the usual on-shell dimensional reduction we have the on-shell relation  
\be \label{varphifixed}
X^I \=\frac{1}{2} (\sigma^I+i\phi W^I) \, \; \Longrightarrow \, \varphi = -\pi/2 \,.
\ee

The 5d supersymmetry transformation upon dimension reduction reduces to a linear combination of 
4d supersymmetry transformation, S-supersymmetry transformation, $SU(2)$-R symmetry transformation, 
and chiral $U(1)$ transformation.
 as 
\be \label{5dto4dsusy}
\delta_{Q}(\epsilon)|^{\text{reduced}}_{5D}\Phi\=\delta_{Q}(\epsilon)|_{4D}\Phi+\delta_{S}(\widetilde\eta)|_{4D}\Phi+\delta_{SU(2)}(\widetilde\Lambda)|_{4D}\Phi+\delta_{U(1)}(\widetilde\Lambda^{0})|_{4D}\Phi\,,
\ee
where the parameters~$\wt \eta$, $\wt \Lambda$, $\wt \Lambda^0$ are non-linear combinations of 
the various supergravity fields~\cite{Banerjee:2011ts}. 
The equation~\eqref{5dto4dsusy} relates supersymmetric configurations in 5d to supersymmetric configurations in 4d.
In bosonic backgrounds, we see that the parameters~$\widetilde\Lambda$ and~$\widetilde\Lambda^{0}$
vanish, and thus the~$\delta_{Q}$ variation in 5d reduces to a combination of~$\delta_{Q}$ and~$\delta_{S}$ in 4d.
Upon demanding the left-hand side of~\eqref{5dto4dsusy} vanishes on the 5d gravitino field~$\Phi$, 
we obtain, from the the right-hand side, precisely the condition for vanishing of the 4d gravitino, as well as a condition
on the KK vector multiplet. 

In our case of interest discussed in Section~\ref{sec:susyBH}, the 5d spinning supersymmetric black hole reduces 
precisely to the 4d supersymmetric black hole with electric and magnetic flux in the KK multiplet. The electric flux is 
proportional to the angular momentum in 5d, while the magnetic flux is the Taub-NUT charge in 5d. The 5d black hole 
sits at the center of the Taub-NUT space. This is simply a restatement, in the off-shell theory, of the 
4d/5d connection~\cite{Gaiotto:2005gf}. 
The advantage of the off-shell formalism is that we can also analyze the off-shell fluctuations relevant for localization. 
For vector multiplet fluctuations around the black hole background, we can check that the parameter~$\widetilde \eta$ also vanishes, and 
the~$\delta_{Q}$ variations in 5d map to the~$\delta_{Q}$ variations in 4d. For Weyl multiplet fluctuations, this is not the 
case and the 5d to 4d reduction necessarily involves the conformal Killing spinor~$\wt \eta$.

However, an important caveat to all these considerations is that these off-shell 5d/4d reductions of~\cite{Banerjee:2011ts} 
are written in Lorentzian space. For the purpose of localization calculations of the functional integral, we are interested 
in Euclidean configurations, for which there may be subtleties in the choice of analytic continuation. For this reason we 
choose a different route and directly analyze the 5d supersymmetry vanishing conditions and explore the choices of 
Euclidean continuation. As we will see in the following sections, these subtleties indeed play an important role.

\section{Off-shell Weyl multiplet analysis \label{sec:Weyl}}

In this section we analyze the off-shell BPS equations in the Weyl multiplet sector described in Section \ref{sec:offshell}. 
We perform our analysis using the spinor bilinear method which yields a set of coupled first order differential equations 
for bosonic quantities. We follow the references~\cite{Tod:1983pm, Gauntlett:2001ur, Gauntlett:2002sc,Bellorin:2005zc,Bellorin:2006yr}.
We then present the most general solution to these equations with some of the auxiliary fields set to zero.

The idea of the spinor bilinear method to find all supersymmetric solutions of a given system is as 
follows. 
We obtain supersymmetric solutions by setting the supersymmetry variations 
of all the fields to zero. Assuming that there are no fermionic backgrounds, we have to set all fermion variations 
to zero, which leads to matrix equations in spinorial variables. Instead of working with these matrix equations,
one begins by assuming the existence of a Killing spinor and forms various bilinears of this spinor. 
The original BPS equations then lead to a set of coupled first order equations for these bosonic 
quantities. These quantities are then interpreted as describing the bosonic background in which one is interested. 
For example, one finds that the vector bilinear obeys the Killing vector equation, and we interpret this to mean 
that the background must have a Killing vector. Carrying on in this manner one finds other constraints on the bosonic 
fields, which we then put together to construct the solution space.

This method has been applied successfully to classify all supersymmetric solutions of various systems. 
Two references that we follow closely in terms of conventions are the classification of on-shell BPS solutions 
of 5d~$\CN=2$ supergravity~\cite{Gauntlett:2002nw}, and the classification of off-shell BPS solutions of 4d~$\CN=2$ 
supergravity~\cite{Gupta:2012cy}. 
From now on we will consider Euclidean configurations.
In Appendix~\ref{Conv.SpinorAlg.}  we present the conventions that we use for Euclidean spinors and gamma matrices. 

\subsection{Killing Spinor and its Bilinears} \label{sec:KS+Bil}

Upon setting the gravitini variations in~\eqref{WeylVar5d} to zero, we obtain the BPS equations:
\be
2\, \mathcal D_{M}  \epsilon^{i}+\frac{i}{2}T_{AB} \, \bigl( 3\g^{AB}\g_{M}-\g_{M}\g^{AB} \bigr) \, \epsilon^{i}-i\g_{M} \, \eta^{i} \= 0 
\ee
for the Killing spinor~$ \epsilon^{i}$ and a similar one for its conjugate~$\overline{\epsilon}_{i}$. 
In our analysis below we set the dilatation gauge field~$b_M$, the~$SU(2)_R$ gauge field~$V_{M j}{}^{i}$, and 
the conformal supersymmetry parameter~$\eta^i$ to zero. We make these assumptions in order to simplify the problem, 
and they need to be revisited in order to have a complete analysis.
With these assumptions the Killing spinor equations take the following form,
\begin{equation} \label{Eqn:KSE}
\begin{split}
\nabla_{M}\epsilon^{i} & \= - \frac{i}{4} \,T_{AB} \, \bigl( 3\gamma^{AB}\gamma_{M} - \gamma_{M}\gamma^{AB} \bigr) \, \epsilon^{i} \,, \\
\nabla_{M}\overline{\epsilon}_{i} & \= +\frac{i}{4} \, T_{AB} \, \overline{\epsilon}_{i}  \, 
\bigl(3\gamma_{M}\gamma^{AB} - \gamma^{AB}\gamma_{M} \bigr) \,.
\end{split}
\end{equation}
Here~$\nabla_M$ is the covariant derivative~$\mathcal{D}_M$ with the above simplifications, and takes the form
\be
\nabla_{M}\epsilon^{i} \= \bigl(\p_{M}-\frac{1}{4}\omega_{M}{}^{AB}\g_{AB} \bigr) \, \eps^{i}  \,.
\ee

Now we start building the spinor bilinears. 
The products of spinor bilinears in $5$ dimensions obey the following conditions~\cite{Gauntlett:2002nw} 
\begin{equation}
\overline{\epsilon} \, \gamma_{M{1}...M_{n}} \,\eta \= - \overline{\eta} \, \gamma_{M_{n}...M_{1}} \, \epsilon \,.
\label{Eqn:flippingspinors}
\end{equation}
In particular, this implies that the product of two Killing spinors, $\overline{\epsilon}^{i}\epsilon^{j}$, is antisymmetric in the 
symplectic indices, $i, j$, and therefore the only non-trivial scalar bilinear is 
\begin{equation}
f \; := \; \frac{1}{2}\varepsilon_{ij} \, \overline{\epsilon}^{i} \, \epsilon^{j} \,. 
\end{equation}
One can check using the reality properties of the spinors that~$f$ is a real scalar. 

Using the BPS equations~\eqref{Eqn:KSE} for the Killing spinor and its conjugate, we see that
\begin{equation}
\begin{split}
\nabla_{A} f &\= \frac{i}{8}\varepsilon_{ij} \, \overline{\epsilon}^{i} \, T^{BC}\bigl(3\gamma_{A}\gamma_{BC} - \gamma_{BC}\gamma_{A} - 3\gamma_{BC}\gamma_{A} + \gamma_{A}\gamma_{BC}\bigr)\epsilon^{j}\,, \\
 &\= \frac{i}{2}\varepsilon_{ij}\overline{\epsilon}^{i}T^{bc}\left[\gamma_{a},\gamma_{bc}\right]\epsilon^{j}\,, \\
&\= 2i \, T^{BC}\varepsilon_{ij} \, \overline{\epsilon}^{i} \, \delta_{A[B}\gamma_{C]}\epsilon^{j}\,,
\end{split}
\end{equation}
where, in going to the second line we have used the gamma matrix identity~\eqref{Eqn:gamma12com}. 
Now using the antisymmetry of $T$, we can rewrite this as
\begin{equation} \label{Eqn:df}
df \= -4i \, \iota_{K}T.
\end{equation}
where $K^{B} = \frac{1}{2}\varepsilon_{ij}\overline{\epsilon}^{i}\gamma^{B}\epsilon^{j}$, and~$\iota_V$ is the interior derivative along~$V$.
Similarly we construct the vector bilinears~$\overline{\epsilon}^{i}\gamma^{A}\epsilon^{j}$. 
Like above, using Equation~\eqref{Eqn:flippingspinors}, we see that this is anti-symmetric in $i,j$, and 
therefore the vector $K$ defined above is the only non-trivial vector bilinear. 
Using the reality properties of the spinors described in Appendix~\ref{Conv.SpinorAlg.}, we see that this is a real vector.

Using the BPS equations, we compute the covariant derivative of $K$,
\begin{equation}
\begin{split}
\ \nabla_{A} K_{B} &\= \frac{1}{2}\varepsilon_{ij}\bigl(\nabla_{A}\overline{\epsilon}^{i}\bigr)
\gamma_{B}\epsilon^{j} + \varepsilon_{ij}\overline{\epsilon}^{i}
\gamma_{A}\bigl(\nabla_{B}\epsilon^{j}\bigr)\,\\
 &\= i\star{T}_{ABC}K^{C}+4ifT_{AB}\,.
 \end{split}
\end{equation}
Since the right hand side of the above equation is antisymmetric in $A, B$, we see that $K$ is a Killing vector,
\begin{equation}
\nabla_{A}K_{B} + \nabla_{B}K_{A} \= 0 \,,
\end{equation}
and that the exterior derivative of the Killing vector $K$ is
\begin{equation} \label{Eqn:dKT}
dK \= 4i\iota_{K}\star{T} + 8iTf \,.
\end{equation}

We now construct two form bilinears $A^{ij}{}_{AB}\equiv \overline{\epsilon}^{i}\gamma_{AB}\epsilon^{j}$. 
Using Equation~\eqref{Eqn:flippingspinors}, we see that these are symmetric in $i,j$. 
The two forms $A^{11}$ and $A^{22}$ are complex conjugates and $A^{12}$ is purely imaginary.
We can describe these complex two forms by the three real two forms~$X_{i}$, $i=1,2,3$, given by
\begin{equation}
A^{11} \= X_{1} + iX_{2} \,, \qquad 
A^{22} \= X_{1} - iX_{2} \,, \qquad 
A^{12} \= -iX_{3} \,.
\end{equation}
Using the Killing spinor equation, the covariant derivative of these two forms can be expressed as
\begin{equation}
\begin{split}
\nabla^{C}A^{ij}{}_{AB} &\= \frac{i}{4}\overline{\epsilon}^{i}T^{DE}\bigl(3\gamma^{C}\gamma_{DE}\gamma_{AB} - \gamma_{DE}\gamma^{C}\gamma_{AB} - 3\gamma_{AB}\gamma_{DE}\gamma^{C} + \gamma_{AB}\gamma^{C}\gamma_{DE}\bigr)\epsilon^{j}\,, \\
&\= \frac{i}{4}\overline{\epsilon}^{i}\bigl(8T^{CD}\bigl\{\gamma_{D},\gamma_{AB}\bigr\} + 2T_{DE} \bigl[\gamma^{DEC},\gamma_{AB} \bigr]\bigr)\epsilon^{j}\,.
\label{Eqn:covtwoform}
\end{split}
\end{equation}
Denoting $C = \star A$ and using gamma matrix identities, we can simplify this to
\be
\ \nabla^{C} A^{ij}{}_{AB} \= 4iT^{CD}C^{ij}{}_{DAB} + 2iT_{A}{}^{e}C^{ij}{}_{ECB} + 2iT_{B}{}^{e}C^{ij}{}_{EAC} +iT^{DE}\bigl(C^{ij}_{DEB}\delta_{AC} - C^{ij}_{DEA}\delta_{BC}\bigr)\,.
\ee
One can check that $\nabla_{[C}A_{AB]} = 0$ so that $A$ is closed, i.e.~$dA = 0$.
Similarly, the covariant derivative of $C$ is
\begin{equation}
\nabla^{D}C^{ij}{}_{ABC} \= 24iT^{D}{}_{[A}A^{ij}{}_{BC]} +18iT_{EF}\delta^{[ED}{}_{[AB}A^{ij}{}^{F]}{}_{C]} \,.
\end{equation}
Noting that $\nabla_{[D}C^{ij}{}_{ABC]} = 18iT_{[DA}A_{BC]}$, we get
\begin{equation}
d\star A \= 18iT\wedge A \,.
\end{equation}

To summarize, we have a real scalar~$f$, a Killing vector~$K$, and the closed two-forms~$A^{ij}$
that obey the differential relations presented in Table \ref{Tab:Genbil}.
\begin{table}[h!]
\centering
\begin{tabular}{ |c|c|c| } 
 \hline
 \textbf{Type} & \textbf{Definition} & \textbf{Derivative} \\
 \hline
 Scalar & $f = \varepsilon_{ij}\overline{\epsilon}^{i}\epsilon^{j}$ & $df = -4i\iota_{K}T$  \\ 
 \hline
 Vector & $K^{A} = \varepsilon_{ij}\overline{\epsilon}^{i}\gamma^{A}\epsilon^{j}$ & $\nabla_{(A}K_{B)} = 0,$ \\ 
 {} & {} & $dK = 2i\iota_{K}\star{T} + 8iTf.$ \\
 \hline
 Two Forms & $A^{ij}_{AB} = \overline{\epsilon}^{i}\gamma_{AB}\epsilon^{j}$ & $dA = 0,$ \\
 {} & {} & $d\star A = 18iT\wedge A.$ \\
\hline
\end{tabular}
\caption{Summary of the differential relations between Killing Spinor Bilinears}
\label{Tab:Genbil}
\end{table}
In fact these fields are not all algebraically independent. We can use the Fierz identities for the products of spinors to 
get algebraic relations between them~\cite{Gauntlett:2002nw}. 
\begin{eqnarray}
\ K^{A}K_{A} &\=& f^{2} \,, \label{Eqn:timenullkill} \\
\ X_{i} \wedge X_{j} &\=& -2\delta_{ij}f\star K \,, \\
\ \iota_{K}X_{i} &\=& 0 \,, \label{Eqn:ikX} \\
\ \iota_{K}\star X_{i} &\=& -f X_{i} \,, \label{Eqn:ikstarX} \\
\ X^{T}_{i}X_{j} &\=& \delta_{ij}\bigl(f^2 \mathbf{1}-K.K^{T}\bigr) + f\varepsilon_{ijk}X^{k} \,, \label{Eqn:XXT}\\
\ K_{A}\gamma^{A}\epsilon^{i} &\=& f\epsilon^{i} \,, \label{Eqn:Kgammaf} \\
\ A^{ij}_{AB}\gamma^{AB}\epsilon^{k} &\=& 8f\varepsilon^{k(i}\epsilon^{j)} \,.
\end{eqnarray}

\subsection{Resulting conditions on the Weyl Multiplet fields}

In the previous subsection, we saw that off-shell supersymmetric metrics in five dimensions possess at least one Killing vector~$K$. 
Since~$f$ is real, Equation~\eqref{Eqn:timenullkill} implies that this Killing vector is either timelike, if $f^2 > 0$, or null, if $f = 0$. 
In this paper we will focus on the timelike case which contains the supersymmetric spinning black holes. The null case also
appears as an extremal spinning limit of these solutions 
(which is in fact~$AdS_{3}\times S^{2}$) as shown in~\cite{Gauntlett:2002nw}, and we leave the analysis of this for future work.

Following the method outlined in~\cite{Gauntlett:2002nw}, we define a coordinate~$\tau$ by $K = \partial_{\tau}$, 
so that~$f^2 = g_{\tau\tau}$.
Expressing the remaining coordinates as $x^{\m}$, $\m =\{1,2,3,4\}$, the metric takes the general form
\begin{equation}
ds^{2} \= f^{2}\bigl(d\tau + \omega\bigr)^{2} + f^{-1}h_{\m\n}dx^{\m}dx^{\n} \,,
\label{Eqn:SpTmetric}
\end{equation}
where $\omega = \omega_{\m}dx^{\m}$ is a one form with $\omega_{\tau} = 0$, 
and $f^{-1}h_{\m\n}$ is the projection of the full metric perpendicular to the orbits of $K$. 
Furthermore, since $K$ is the Killing vector, the components of the metric are independent of the coordinate $\tau$, i.e.
\be
\p_{\tau}f(x)\=0\,,\quad \p_{\tau}\omega_{\m}(x)\=0\,,\quad \p_{\tau}h_{\m\n}(x)\=0\,.
\ee
We view this as a fibration of the~$\tau$ coordinate over the base-space~$x^m$. As we now explain, the 
relations obtained in the previous subsection imply that the tensor~$T$, and therefore the full field configuration, 
is determined in terms of the parameters~$f$, $\omega$, and~$h_{mn}$, and, further,~$h_{mn}$
is constrained to be hyper-K\"ahler. This is exactly as in the on-shell analysis, with the auxiliary tensor~$T$
in the off-shell theory playing the role of the graviphoton field strength~$F$ in the on-shell theory.

We begin by writing $K=fe^{1}$, where $e^{1} = f\big(d\tau + \omega\big)$. Its exterior derivative obeys
\begin{equation} \label{dKfdOm}
dK = d\bigl(fe^{1}\bigr) \= 2df \wedge e^{1} + f^{2}d\omega \,.
\end{equation}
We denote the self-dual and anti-self-dual components of~$fd\omega$ with respect to the base metric~$h$ by $G^{+}$ and $G^{-}$.  
Comparing the two expressions for the exterior derivative of $K$ in Equations~\eqref{dKfdOm} and~\eqref{Eqn:dKT}, we obtain
\begin{equation}\label{Eqn:Tnotcomp}
\ T + \frac{1}{2}f^{-1}\iota_{K}\star T \= -\frac{i}{8}\bigl(-2f^{-2}K\wedge df + G^{+} + G^{-}\bigr)\,.
\end{equation}
We can now solve  for~$T$. It is useful to write the above equation in component form:
\begin{equation} \label{Eqn:Tcomp}
T_{AB} \+ \frac{1}{4}\varepsilon_{1ABCD}T^{CD} \= \frac{i}{8}\bigl(2f^{-2}K_{[A}\nabla_{B]}f - G^{+}_{AB} - G^{-}_{AB}\bigr)\,.
\end{equation}
The mixed components between the base space and fibre, i.e.~$A = 1, B = b$ of  Equation~\eqref{Eqn:Tcomp} is
\begin{equation}
T_{1b} \= \frac{i}{8}\bigl(2f^{-1}\nabla_{b}f - G^{+}_{1b}- G^{-}_{1b}\bigr)\,.
\end{equation}
The fact that $fd\omega = G^{+} + G^{-}$ lives on the base space implies that its projection~$G^{+}_{1b} + G^{-}_{1b} = 0$, and therefore 
\begin{equation}
T_{1b} \= \frac{i}{4}f^{-1}\nabla_{b}f.
\label{Eqn:Tcomp0m}
\end{equation}
The $A = a, B = b$ components of  Equation~\eqref{Eqn:Tcomp}, using~$K_a=0$, is
\be
T_{ab} \+ \frac{1}{4}\varepsilon_{abcd} \, T^{cd} \= -\frac{i}{8}\bigl(G^{+}_{ab} + G^{-}_{ab}\bigr) \,,
\ee
where $\varepsilon_{abcd}$ is the 4d Levi Civita tensor. Multiplying the above equation with $\varepsilon^{abcd}$, 
we obtain its 4d Hodge dual. Putting these two together we obtain
\begin{equation}
T_{ab} \= -\frac{i}{12}\bigl(G^{+}_{ab} + 3G^{-}_{ab}\bigr)\,.
\end{equation}
Thus we have obtained all the components of~$T$, which can be summarized as
\begin{equation}
T \= -\frac{i}{4} de^{1} + \frac{i}{6}G^{+} \,.
\label{Eqn:Tfinal}
\end{equation}

We now look for conditions on the two forms and the spatial metric~$h_{mn}$. 
Equation~\eqref{Eqn:ikX} implies
\begin{equation}
K^{a}X^{i}_{ab} \= fX^{i}_{1b} \= 0 \,,
\end{equation}
and therefore the only non-zero components of the two forms~$X^i$ are on the base space.
Now Equation~\eqref{Eqn:ikstarX} implies that the two forms are anti-self dual on the base space, i.e.,
\be 
\star_{4}X^{i} \= -X^{i}.
\ee
If we consider the two forms to live on the space with $h_{mn}$ 
rather than $f^{-1}h_{mn}$, we must shift them by $X^{i}\rightarrow -fX^{i}$, 
where the negative sign is added for convenience. The algebraic relation given by Equation~\eqref{Eqn:XXT} simplifies to
\begin{equation}
X^{i}_{mp}X^{j}{}^{pn} \= -\delta_{ij}\delta_{m}^{n} + \varepsilon_{ijk}X^{k}_{m}{}^{n} \,.
\end{equation}
This shows that the two forms satisfy a quaternionic algebra over the base space. 
The fact that the two forms~$X^i$ are closed then implies that the metric~$h_{mn}$ is 
integrable and hyper-K\"ahler.

\vspace{0.4cm}

To summarize, we have shown that the most general solution for the metric is
\begin{equation}
ds^{2} \= f^{2}\bigl(d\tau+\omega\bigr)^2 + f^{-1}h_{mn}dx^{m}dx^{n} \,,
\label{Eqn:finalmetric}
\end{equation}
where $\omega = \omega_{m}dx^{m}$ is a one form in the spatial direction, 
and $h$ is a hyper-K\"ahler four dimensional manifold. 
The coordinate $\tau$ is defined by the vector $K\equiv \frac{1}{2}\varepsilon_{ij}\epsilon^{i}\gamma\epsilon^{j} = \partial_{\tau}$. 
The scalar~$f \equiv \frac{1}{2}\varepsilon_{ij}\epsilon^{i}\epsilon^{j}$ satisfies $f^{2}=K^{\mu}K_{\mu}$. 
As $K$ is a Killing vector, $f, \omega$ and $h$ are independent of $\tau$.
The $T$ field is given by
\begin{equation}
T \= -\frac{i}{4}de^{1} + \frac{i}{6}G^{+},
\label{Eqn:finalT}
\end{equation}
where $e^{1} = f\bigl(d\tau+\omega\bigr)$ and $G^{\pm}$ are the self dual and anti-self dual parts of~$fd\omega$ with respect to the spatial metric.

\vspace{2mm}

In Section~\ref{sec:susyBH} we have seen that the near-horizon configuration of rotating supersymmetric black holes in 5d $\CN=2$
supergravity are a maximally supersymmetric solution. Therefore, in particular, it should belong to the set of 
general backgrounds described by~\eqref{Eqn:finalmetric} and~\eqref{Eqn:finalT}. We now proceed to verify this assertion. 
The near-horizon limit of our rotating black holes is the $AdS_{2}\times S^{2} \ltimes S^1$ metric~\eqref{NearHorzMetric}. 
We analytically continue this metric to Euclidean space by taking~$\beta \rightarrow i\a$ with real~$\a$, as in~\cite{Gomes:2013cca}. 
We will sometimes refer to the parameter~$\a$ as the rotation parameter of the black hole.
We also set the constant~$e^{g} = \cosh\alpha$, so that the Euclidean metric is 
\begin{equation}
ds^{2} \= r^{4}d\tau^{2} + 4r^{-2}dr^{2} + d\psi^{2} + \sin^{2}\psi d\phi^{2} + \bigl(\cosh\alpha\bigl(d\rho + \cos\psi d\phi\bigr)
- \sinh\alpha r^2 d\tau\bigr)^{2} \,.
\label{Eqn:AdS2S3Euc}
\end{equation}
The auxiliary tensor~$T$, given by~\eqref{NearHorzT}, is now analytically continued by 
taking~$T_{01} \rightarrow iT_{12}$, so that we have
\be \label{Eqn:Teucl}
T_{12} \= -\frac{i}{4} \, \cosh \a \,, \qquad T_{34} \= \frac{i}{4} \, \sinh \a \,.
\ee
This metric can be rewritten as 
\begin{equation}
ds^{2} \= r^{4}\cosh^{2}\alpha\bigl(d\tau -r^{-2}\tanh\alpha\bigl(d\rho + \cos\psi d\phi\bigr)\bigr)^{2} + \frac{1}{r^{2}\cosh\alpha} h \,,
\label{Eqn:AdS2S3r}
\end{equation}
where
\begin{equation} \label{basespace}
h \= \cosh\alpha\Bigl(4dr^{2} + r^{2}\bigl(d\psi^{2} + \sin^{2}\psi d\phi^{2} + \bigl(d\rho + \cos\psi d\phi\bigr)^2\bigr)\Bigr) \,.
\end{equation}
Thus we can identify the near-horizon rotating black hole metric with~\eqref{Eqn:finalmetric} with
\be
f \=r^2\cosh\alpha \,, \qquad \omega \= -r^{-2}\tanh\alpha\bigl(d\rho + \cos\psi d\phi\bigr) \,.
\ee 
With these identifications, the self dual part of~$fd\omega$ with respect to the base-space metric~$h$ 
is given by $G^{+} = 3\sinh\alpha\bigl(2e^2 \wedge e^5 + e^3 \wedge e^4\bigr)$. 
Equation~\eqref{Eqn:finalT} then yields the value of $T$ for the metric given by \eqref{Eqn:AdS2S3r} 
\be
\ T \= -\frac{i}{4} de^{1} + \frac{i}{6}G^{+} \= -\frac{i}{2}e^1 \wedge e^2 + \frac{i}{4}\sinh\alpha e^{3}\wedge e^{4} \,,
\label{Eqn:TAdS2S3}
\ee
where, as before, $e^{1} = f\bigl(d\tau+\omega\bigr)$.
From a simple change of coordinates one sees that the base space given by \eqref{basespace} is $\mathbb{R}^{4}$, exactly as 
in~\cite{Gauntlett:2002nw}, which is hyper-K\"ahler. 

The complete localization analysis involves finding all solutions in the Weyl multiplet sector with boundary conditions 
set by the above fully supersymmetric near-horizon configuration. As mentioned in the introduction this is a difficult problem
that we postpone to future work.\footnote{The analogous problem in the on-shell theory with asymptotically flat boundary 
conditions has been solved in~\cite{Kunduri:2014iga}. It would be interesting if these methods can be generalized
to the off-shell case.} In the following section we will fix the Weyl multiplet to be this near-horizon black hole 
configuration, and calculate the most general supersymmetric off-shell fluctuation of the vector multiplet around this background.

\section{Off-shell vector multiplet analysis \label{sec:Vector}} 

In this section we classify the complete set of off-shell BPS solutions in the vector multiplet sector. We begin with an algebraic 
analysis of the vector multiplet fluctuations around a general background, and then apply this analysis to our case 
of interest, namely the supersymmetric black hole solution discussed in Section~\ref{sec:susyBH}. We pay close 
attention to the analytic continuation of the Euclidean fields and discuss different choices. 

We analyze the BPS vector multiplet fluctuations around a given Weyl multiplet BPS background 
with Killing spinor~$\eps^i$. We have to analyze 
the vanishing of the supersymmetry variations of the gaugini given in Equation~\eqref{VectorSusy}.
Putting the fermionic backgrounds to vanish, we obtain
\begin{equation} \label{BPSvector}
\bigl(F^{AB}-4\sigma T^{AB} \bigr)\gamma_{AB}\epsilon^{i} + 2i\partial_{M}\sigma \gamma^{M}\epsilon^{i} + 4\varepsilon_{jk}Y^{ij}\epsilon^{k} \= 0 \,,
\end{equation}
where $F_{AB} = 2\partial_{[A}W_{B]}$. We recall that we have set the superconformal transformation parameter $\eta^{i}$ to zero 
as discussed in the Section~\eqref{sec:KS+Bil}.
In Section~\ref{sec:susyBH} we presented the attractor near-horizon solution for the spinning black hole. 
The fluctuations of the fields of a given vector multiplet are defined as an expansion around their 
attractor values~\eqref{NearHorzGaugeFld}
\be
F^{AB} \= F^{AB}_* + f^{AB} \,, \qquad \sigma \= \sigma_* + \Sigma \,.
\ee
The attractor value for the auxiliary~$Y^{ij}$ is zero, and we shall continue to denote its fluctuation by the same name.
Our task now is to find all solutions to Equation~\eqref{BPSvector} for all vector multiplet field fluctuations with vanishing asymptotic values.

\subsection{Algebraic analysis}

We start with Equation~\eqref{BPSvector} that our chosen Killing spinor satisfies, and multiply it by the conjugate Killing spinor.
The equations now contain the spinor bilinears that we defined in Section~\ref{sec:Weyl}, namely the scalar~$f$, the Killing vector~$K$,
and the two-forms~$A^{ij}$. Multiplying the BPS equation~\eqref{BPSvector} for~$i=1,2,$ by the conjugate spinors~$\overline{\eps}^{2}$, 
$\overline{\eps}^{1}$, respectively, we obtain four equations involving these bilinears and the fluctuations of the bosonic fields. 
After some rearrangement these four equations can be expressed as follows, 
\begin{equation} \label{EqnsSetA}
\begin{split}
K^{M}\partial_{M}\Sigma & \= 0 \,, \\
\bigl(f^{AB}-4\Sigma T^{AB}\bigr)A^{12}_{AB} + 4Y^{12}f & \= 0 \,, \\
\ \bigl(f^{AB}-4\Sigma T^{AB}\bigr)X^{1}_{AB} - 2\bigl(Y^{11} + Y^{22}\bigr)f &\= 0 \,,    \\
\ \bigl(f^{AB}-4\Sigma T^{AB}\bigr)iX^{2}_{AB} + 2i\bigl(Y^{11} - Y^{22}\bigr)f &\= 0 \,. 
\end{split}
\end{equation}
Similarly, upon multiplying the BPS equation~\eqref{BPSvector} for~$i=1,2,$ by the conjugate spinors~$\overline{\eps}^{2} \gamma^C$, 
$\overline{\eps}^{1} \gamma^C$, respectively, for~$C=1,\dots,5$, and after some rearrangement, we obtain the following equations 
\be \label{EqnsSetB} 
\ \bigl(f_{CB}-4T_{CB}\Sigma\bigr)k^{B}+ i\partial_{C}\Sigma f \= 0 \,, \qquad c=1,\dots,5 \,.
\ee 
One of these five equations is actually implied by~\eqref{EqnsSetA} so that we only have four independent equations.
We recall from Section~\ref{sec:Weyl} that the bilinears $f$ and $K$ are real, and the auxiliary field $T$ is imaginary. 
Taking the fluctuations $\Sigma$ and $f_{CB}$ to be real, we obtain 
\begin{equation} \label{Eqn:Kf}
K^B f_{CB}=0
\end{equation}

Thus we reach the equations~\eqref{EqnsSetA},~\eqref{EqnsSetB}, which 
hold for any background which admits a Killing spinor~$\eps^i$. 
In addition to these eight equations we also impose the Bianchi identities as usual. 
Given a Weyl multiplet background, the analysis reduces to 
computing the Killing spinor bilinears defined in Section~\ref{sec:Weyl}, and finding the 
most general solutions to the above eight equations and the Bianchi identities. 

The analysis so far has been quite general and may be useful to analyze 
vector multiplet fluctuations in a wide variety of circumstances in the context of theories with eight supercharges. 
Now we move to our case of interest, namely the near-horizon region of the 5d black hole,
which is given in~\eqref{Eqn:AdS2S3Euc}. It is convenient to use a different set of coordinates, as in \citep{Gomes:2013cca},
in which the near-horizon region has the following form,
\be \label{Eqn:AdS2S2S1euc}
\begin{split}
ds^{2} & \= \sinh^{2} \eta d\theta^{2} + d\eta^{2} + d\psi^{2} + \sin^{2}\psi d\phi^{2} + \cosh^{2}\alpha\bigl(d\rho + B\bigr)^{2}\,, \\
B & \= + \cos\psi d\phi - \tanh\alpha(\cosh\eta - 1)d\theta\,.
\end{split}
\ee
This configuration admits eight Killing spinors which we present in Appendix \ref{App:KS}. 
For the localization computation we need one Killing spinor which we choose to be
\be \label{KillSpinChosen}
\epsilon^{1}\=e^{\frac{i}{2}(\theta+\phi)}
\begin{pmatrix}
e^{-\frac{\eta}{2}}\cos\frac{\psi}{2}(e^{\eta}\sinh\frac{\a}{2}+\cosh\frac{\a}{2})\\
ie^{-\frac{\eta}{2}}\sin\frac{\psi}{2}(e^{\eta}\sinh\frac{\a}{2}-\cosh\frac{\a}{2})\\
-ie^{-\frac{\eta}{2}}\cos\frac{\psi}{2}(e^{\eta}\cosh\frac{\a}{2}+\sinh\frac{\a}{2})\\
e^{-\frac{\eta}{2}}\sin\frac{\psi}{2}(e^{\eta}\cosh\frac{\a}{2}-\sinh\frac{\a}{2})
\end{pmatrix}.
\ee

We will begin with the case of the static black hole $(\alpha=0)$ which is simpler, and then move on to the case of 
the spinning black hole. 

\subsection{Vector multiplet fluctuations around $AdS_{2}\times S^{3}$ (static black hole)}

The bilinears corresponding to our Killing spinor~\eqref{KillSpinChosen} with~$\a=0$ are summarized in Table~\ref{Tab:Statbil}.
\begin{table}[h]
\centering
\begin{tabular}{ |c|c| } 
 \hline
 \textbf{ } & \textbf{Static black hole:  Killing spinor Bilinears} \\ \hline
 $f$ & $f = -4\cosh\eta$  \\ 
 \hline
 $K^{A}$ & $K^{1} = 4\sinh\eta, \quad K^{4} = -4\sin\psi,\quad K^{5} = -4\cos\psi ,$ \\ 
 \hline 
 $A^{12}_{AB}$ & 
$A^{21}_{12} = -2i , \quad A^{21}_{24} = 2i\sin\psi\sinh\eta , \quad A^{21}_{25} = 2i\cos\psi\sinh\eta ,$ \\
{} & $A^{21}_{34} = 2i\cos\psi\cosh\eta , \quad A^{21}_{35} = -2i\sin\psi\cosh\eta ,$ \\
 \hline 
& $A^{11}_{13} = 2e^{i(\theta+\phi)} , \quad A^{11}_{14} = 2ie^{i(\theta+\phi)}\cos\psi , \quad A^{11}_{15} = -2ie^{i(\theta+\phi)}\sin\psi ,$ \\
 $A^{11}_{AB}$ & $A^{11}_{23} = -2ie^{i(\theta+\phi)}\cosh\eta , \quad A^{11}_{24} = 2e^{i(\theta+\phi)}\cos\psi\cosh\eta , \quad A^{11}_{25} = -2e^{i(\theta+\phi)}\sin\psi\cosh\eta,$ \\ 
{} & $A^{11}_{34} = -2e^{i(\theta+\phi)}\sin\psi\sinh\eta , \quad A^{11}_{35} = -2e^{i(\theta+\phi)}\cos\psi\sinh\eta , \quad A^{11}_{45} = -2ie^{i(\theta+\phi)}\sinh\eta ,$ \\
\hline
\end{tabular}
\caption{Independent non-zero spinor bilinears for the Killing spinor, $\epsilon^{1}_{++}$ of $AdS^{2}\times S_{3}$.}
\label{Tab:Statbil}
\end{table}

We now write the eight basic BPS equations~\eqref{EqnsSetA},~\eqref{EqnsSetB} in this context. 
The equations~\eqref{EqnsSetA} are
\bea
\bigl(2k_{3}\cosh\eta - \Sigma\bigr)i &\= &
\frac{1}{\sinh\eta}\bigl(f_{\theta\eta} - f_{\eta \phi}\sinh^2\eta\bigr) +\frac{\cosh\eta}{\sin\psi} \bigl(f_{\psi \rho} - f_{\psi\phi} \cos\psi\bigr)\,, \label{Eqn:SPT2A1} \\
k_{+} \, \cosh\eta &\= &
\frac{1}{\sinh\eta}\bigl(f_{\theta\psi} - f_{\psi\phi}\sinh^2\eta\bigr) -
\frac{\cosh\eta}{\sin\psi} \bigl(f_{\eta\rho} - f_{\eta\phi}\cos\psi\bigr)\,, \label{Eqn:SPT2A2}  \\
i k_{-} \cosh\eta &\= & 
f_{\eta\psi}\cosh\eta +\frac{1}{\sinh\eta\sin\psi}\bigl(f_{\theta\rho}-f_{\theta\phi} \cos\psi\bigr)+f_{\phi\rho}\frac{\sinh\eta}{\sin\psi}  \,, \label{Eqn:SPT2A3}
\eea
and the equations~\eqref{EqnsSetB} become the following five equations for $M = (\theta, \eta, \psi, \phi, \rho)$,
\be
-i\p_{M}\bigl(\Sigma\cosh\eta\bigr) \= f_{\theta M} + f_{M\phi}\,.
\label{Eqn:SPT2B}
\ee
Here we have set~$Y^{12}=k^3$, $Y^{11} = k_{1}e^{i(\theta + \phi)}, Y^{22} = k_{2}e^{-i (\theta + \phi)}$,  
and $k_{\pm}=\frac12 (k_{1}\pm k_{2})$.\footnote{We note that the fields~$Y^{11}$ and~$Y^{22}$ are set to zero by hand
in the corresponding treatment in \citep{Gomes:2013cca}.} This is a choice of reality condition, which is the 
same choice made in~\cite{Dabholkar:2010uh,Gupta:2012cy} for the corresponding 4d problem. As noted there, the condition changes as a function 
of the spacetime point, and is set up so that the auxiliary fields~$Y^{11}, Y^{22}$ have the same phase as the Killing 
spinor bilinears $A^{11}$ and $A^{22} = (A^{11})^{*}$ presented in Table \ref{Tab:Statbil}. 

\subsection*{Reality conditions}

Now we turn to an important topic, namely the reality conditions. In the above analysis we already made some choices of 
reality conditions consistent with the 4d analysis of~\cite{Dabholkar:2010uh,Gupta:2012cy}. In our current 5d problem we also 
have to decide the reality properties of the fifth component of the gauge field.
As we explained in the introduction, we will make some choices which look reasonable, and explore their consequences. 
For a given choice each of the equations~\eqref{Eqn:SPT2A3}--\eqref{Eqn:SPT2B} will split into 
real and imaginary parts, and we analyze these two parts separately. As we shall see, the solution set depends on this split quite strongly. 
Further the static and spinning cases seem to differ at this point---this is not surprising given 
that spinning black hole metrics continued to Euclidean space naturally introduce complex metrics. 

As we briefly discussed in Section~\ref{4d5dOffShell}, the fifth component of the gauge field~$W_\rho \= \phi W_5$ 
plays a special role in that it combines with the scalar~$\s$ to form the scalars~$X$, $\overline{X}$ in 4d upon 
dimensional reduction.  
In particular, the dictionary~\eqref{4d5dvec}, \eqref{varphifixed} 
between 4d and 5d vector multiplets for the attractor configuration implies that
\be \label{XWrhorel}
X \= \frac{1}{2}(\sigma+iW_\rho) \,, \qquad \overline X \= \frac{1}{2}(\sigma-iW_\rho) \,.
\ee
In the 4d problem, the fluctuations of~$X$, $\overline{X}$ around their respective attractor values~$X_*$, $\overline X_*$ 
was split, in the Euclidean theory, as
\be \label{HJchoice}
X - X_* \= H + J \,, \qquad \overline X - \overline X_* \= H - J \,.
\ee
In the 4d Lorentzian theory the scalars~$X$ and $\overline X $ are complex conjugate and therefore the corresponding 
fluctuations are~$H+iJ$, $H-iJ$. The choice~\eqref{HJchoice} was made in~\cite{Dabholkar:2010uh,Gupta:2012cy} 
in order to obtain sensible localization solutions, this was later justified in~\cite{deWit:2017} by a more formal treatment of 4d Euclidean supergravity. 

Lifting this to 5d naturally leads us to~$W_\rho$ being purely imaginary, instead of purely real as in the 5d Lorentzian theory. 
We shall call this choice~A1. In fact we have to be a little more precise. 
In addition to Equation~\eqref{XWrhorel} that relates~$W_\rho$ to the scalars in four dimensions,  
the dimensional reduction formula relates the rest of the gauge field components in five and four dimensions as follows,
\be \label{Eqn:W4d}
W_{\mu} \= A_{\mu} + B_{\mu}W_{\rho}\,, \quad \mu = \theta, \eta, \psi, \phi \,,
\ee
In the choice of analytic continuation, we also have to specify the reality conditions for these gauge field components. 
The choice A1 is completely specified by demanding that $W_{\rho}$ is imaginary and that the $4d$ gauge fields $A_{\mu}$ are real.

The choice A1 is consistent with the reduction to four dimensions. 
Another analytic continuation would be to chose $W_{\rho}$ imaginary and $W_{\mu}$ real. We call this A2. 
A third choice would be to simply take all fields to be real, this could be called natural from a purely five-dimensional perspective. 
We call this choice B.
A fourth choice is to look at only the field equations, in this case the supersymmetry equations instead of the fundamental 
variables in the functional integral. This would mean prescribing reality conditions for the field strengths of all thes gauge fields. 
We explore the condition~$f_{\mu \rho}$ imaginary which is related to A1 in that it singles out the direction~$\rho$. 
We call this choice C.

We now present the results for all the reality conditions. 
We have to solve for field variables~$f_{ab}$, $k_{\pm}$, $k^3$, and $\Sigma$. 
In all four cases we find that $\Sigma = C\, \text{sech} \eta$, $k_{3} = (C/2) \, \text{sech}^{2} \eta$, and~$k_- =0$. 
When reduced to 4d, these are precisely the solutions found in~\cite{Dabholkar:2010uh,Gupta:2012cy}. 
The other fields depend on the analytic continuation. 
For choices A1, A2, and C, we find that in fact all the other field fluctuations vanish. 
This degeneracy among the choices is specific to the static black hole and, as we will see in the next section, 
it will be lifted in the spinning black hole. 
For choice B, i.e.~when~$W_\rho$ is real, we are left with the following
constraint that relates the field strengths to~$k_+$,
\begin{equation} \label{Eqn:fconstraint}
f_{MN} - \frac{1}{2}\varepsilon_{MNRST}f^{RS}K^{T} \= k_{+}f^{-1}\text{Re}\bigl(e^{-i(\theta + \phi )}A^{11}_{MN}\bigr)\,.
\end{equation}
The solution for all the fields in the vector multiplet for all our choices of analytic continuations 
are summarised in Table \ref{Tab:Statsol}. 
\begin{table}[h]
\centering
\begin{tabular}{ |c|c|c|c|c| } 
 \hline
& \textbf{A1}  & \textbf{A2}  & \textbf{B}  & \textbf{C} \\[-2mm]
 & {$W_{\rho}$ Im, $A_\mu$ Real} & {$W_{\rho}$ Im}, $W_\mu$ Real & {$W_{\rho}$ Real}  & {$f_{\mu\rho}$ Im} \\
 \hline
 \hline
 $\Sigma$ & $C \ \text{sech}\eta$ & $C \ \text{sech}\eta$ & $C \ \text{sech}\eta$ &$C \ \text{sech}\eta$ \\
 \hline
 $W_{\rho}$ & $0$ &  $0 $ & Solutions to \eqref{Eqn:fconstraint} & $0$ \\
 \hline
 $\wt{f}_{\mu\nu}$, $k_+$ & 0, 0  & 0, 0 & Solutions to \eqref{Eqn:fconstraint} & 0, 0\\ 
 \hline
   $k_{-}$ & 0 & 0 & 0 & 0 \\ 
 \hline 
 $k_{3}$ & $(C/2) \ \text{sech}^2\eta$ & $(C/2) \ \text{sech}^2\eta$ & $(C/2) \ \text{sech}^2\eta$ & $(C/2) \ \text{sech}^2\eta$ \\ 
\hline
\end{tabular}
\caption{The complete set of off-shell BPS solutions for vector multiplet fluctuations around the near-horizon 
static black hole for different reality conditions. Here~$f_{MN}$ is the fluctuation of the field strength of~$W_M$, 
$\wt f_{\m\n}$ is the fluctuation of the field strength of~$A_{\m}$, $\Sigma$ is the 
fluctuation of the scalar~$\sigma$, and~$k_\pm, k^3$ are the fluctuations of the auxiliary fields~$Y^{ij}$. 
For condition~B, some explicit solutions to Equations \eqref{Eqn:fconstraint} are presented in~\cite{Gupta:2015gga}.}
\label{Tab:Statsol}
\end{table}

Now we present the details in each of the cases. 

\vspace{0.2cm}

\subsubsection*{Reality Condition A1}
The reality condition A1 is that $W_{\rho}$ is purely imaginary, the $4d$ gauge fields~$A_\mu$ are real, and 
the remaining fields~$\Sigma, k_{\pm}, k_{3}$ are real.
We recall that the boundary conditions of our problem impose that all the field fluctuations should 
vanish asympotitcally as $\eta \rightarrow \infty$. 
We decompose the gauge fields according to \eqref{Eqn:W4d} and use the value \eqref{Eqn:AdS2S2S1euc} for the 
background fields, with $\alpha = 0$,  $B = \cos\psi \, d\phi$, so that
\be
W_{\theta} \= A_{\theta} \,, \quad  
W_{\eta} \= A_{\eta} \,,\quad
W_{\psi} \= A_{\psi} \,,    \quad
W_{\phi} \= A_{\phi} + \cos\psi W_{\rho}\,.  
\ee
The imaginary parts of the second set of complex equations~\eqref{Eqn:SPT2B} now reduce to
\be \label{Eqn:Sigmaw5}
\Sigma\cosh\eta - iW_{\rho}\cos\psi \= C\,,
\ee
where $C$ is a constant. From the imaginary part of \eqref{Eqn:SPT2A2}, we get
\be \label{Eqn:Wrhodiff}
(\coth\eta \, \p_{\eta} + \cot\psi \,\p_{\psi} - 1)W_{\rho} \= 0\,.
\ee
This equation was analyzed in~\citep{Gupta:2012cy} in the context of 4d theories with~$AdS_2 \times S^2$ boundary conditions,
and it was shown that there are no smooth non-zero solutions to this equation that respect the boundary conditions, as we now briefly recall.
The structure of the differential operator on the left-hand side of~\eqref{Eqn:Wrhodiff} implies that any solution has the form
\be
W_{\rho} \= f\bigl(v, \theta,\phi, \rho\bigr)\sqrt{\frac{\cosh\eta}{\cos\psi}}\,,
\ee
where $f$ is an arbitrary function and $v = \cosh\eta\cos\psi$. 
Writing this as a power series in the variable~$v$, we see that the boundary conditions imply that we only have 
negative powers of~$\cosh\eta\cos\psi$, which is then singular at the points where~$\cos \psi$ vanishes. 
Thus the boundary conditions and smoothness conditions imply that
\be 
W_{\rho} \= 0\,, 
\ee
and, by \eqref{Eqn:Sigmaw5},
\be \label{Eqn:4dresultsigma}
\Sigma \= \frac{C}{\cosh\eta}\,.
\ee
From the imaginary parts of~\eqref{Eqn:SPT2A3} and~\eqref{Eqn:SPT2A1}, we get $k_{-} = 0 $ and
\be
k_{3} \= \frac{C}{2\cosh^2 \eta}\,.
\label{Eqn:4dresultk3}
\ee
The condition $W_{\rho} = 0$ implies that $W_{\mu} = A_{\mu}$. The real parts of the basic equations 
\eqref{Eqn:SPT2A1}--\eqref{Eqn:SPT2B} lead to the following constraints on the field strengths and $k_{+}$,
\begin{equation}
\begin{split}
\ f_{\theta\eta} & \= f_{\phi\eta} = \frac{\bigl(k_{+}\sin\psi - \partial_{\rho}W_{\eta}\bigr)\sinh^{2}\eta\, 
\sin 2\psi - \partial_{\rho}W_{\psi}\sinh 2\eta\sin^{2}\psi}{\bigl(\cos 2\psi -\cosh 2\eta\bigr)\sin\psi}\,, \\
\ f_{\theta\psi} & \= f_{\phi\psi} = -\frac{\bigl(k_{+}\sin\psi - \partial_{\rho}W_{\eta}\bigr)\sinh 2\eta\, \sin^{2}\psi + 
\partial_{\rho}W_{\psi}\sinh^{2}\eta\sin 2\psi}{\bigl(\cos 2\psi -\cosh 2\eta\bigr)\sin\psi}\,,  \\
\ f_{\theta\phi} & \= 0\,, \\
\ f_{\eta\psi} & \= \frac{\partial_{\rho}W_{\theta}}{\tanh\eta\sin\psi}\,,\\
\  \partial_{\rho}W_{\theta} & \= \partial_{\rho}W_{\phi} \,.
\label{Eqn:fcow5}
\end{split}
\end{equation}

\vspace{2cm}

The first thing we note about these equations is that they have a symmetry in~$\theta \leftrightarrow \phi$, 
which implies that the various fields are actually functions of~$\theta+\phi$ only. 
Now we make a gauge choice. For a periodic variable such as~$\theta$ it is not possible to bring a~$\theta$-independent
configuration of~$W_\theta$ to zero by a gauge transformation that respects the periodicity. Therefore we choose 
the gauge condition~$W_\psi=0$ (recall that~$\psi \in [0,\pi]$ is not a periodic variable).\footnote{A similar analysis 
can be done in the gauge~$W_\eta =0$ which also leads to the same final conclusion of no non-trivial smooth solutions
to the system of equations~\eqref{Eqn:fcow5}.}

Next we use the Bianchi identity  
\be
\p_\theta \, f_{\eta\psi} + \p_\eta \, f_{\psi\theta} + \p_\psi \, f_{\theta\eta}  \= 0 \,.
\ee
This equation together with the BPS equations~\eqref{Eqn:fcow5} 
can be rearranged to obtain:
\be
\frac{1}{\tanh \eta \sin \psi} \, \p_\theta \p_\rho W_\theta + \p_\eta \p_\psi W_\theta + \tanh\eta \, \p_\psi \bigl( \cot\psi \, \p_\psi W_\theta \bigr) \=0 \,.
\ee
We now show that this equation has no smooth solution for~$W_\theta$.
In order to do so, we write the field~$W_\theta$ in Fourier components 
\be \label{Wthetaeqn}
W_{\theta}(\theta, \phi, \eta, \psi, \rho)  \= \sum_{p,q \= -\infty}^{\infty} e^{ip(\theta+\phi)}e^{iq\rho} \, W^{(p,q)}_{\theta} (\eta, \psi) \,,
\ee
where we have used the symmetry~$\theta \leftrightarrow \phi$ mentioned above.

The~$q = 0$ mode for each field, i.e.~the modes independent of $\rho$ reduce to the equivalent equations in $4d$.
This 4d problem was analyzed in~\cite{Gupta:2012cy}, using the method that we discussed above~\eqref{Eqn:Wrhodiff}, 
and the conclusion is that there are no non-zero smooth solutions which 
respect the boundary conditions. (We review this again while discussing Condition~C.)
Thus we set~$q\neq 0$ in the following.
The equation~\eqref{Wthetaeqn} now reads
\be
-\frac{pq}{\tanh \eta \sin \psi} \,  W^{(p,q)}_{\theta} (\eta, \psi) 
+ \p_\eta \p_\psi  W^{(p,q)}_{\theta} (\eta, \psi)+ \tanh\eta \, \p_\psi \bigl( \cot\psi \, \p_\psi W^{(p,q)}_{\theta} (\eta, \psi) \bigr) \=0 \,.
\ee
After clearing denominators we have
\be \label{Wpqeqn}
\Bigl( - pq  \, \sin \psi +   \tanh\eta \sin^2 \psi \p_\eta \p_\psi   + \tanh^2\eta \, \bigl(\sin \psi \cos \psi \p^2_\psi -\p_\psi \bigr) \Bigr)
W^{(p,q)}_{\theta} (\eta, \psi) \=0 \,.
\ee

First we consider the~$p=0$ mode, for which this equation simplifies to
\be
\coth\eta\, \p_{\eta}\p_{\psi}W_{\theta} \= -\p_{\psi}(\cot\psi \, \p_{\psi}W_{\theta})\,.
\ee
We can solve this equation by separation of variables, $\p_{\psi}W_{\theta} = f(\eta)g(\psi)$, to obtain
\be
\frac{1}{f}\coth\eta \, \p_{\eta}f \= -\frac{1}{g}\p_{\psi}(g\cot\psi) \= C\,,
\ee
where $C$ is independent of $\eta, \psi$. Solving this
\be
f(\eta) \= A(\cosh\eta)^C\,, \quad g(\psi) \= B\sin\psi (\cos\psi)^{C-1}\,,
\ee
where $A, B$ are independent of $\eta, \psi$. 
When~$C=0$ we see that~$g(\psi)$ (and therefore~$W_\theta)$ is singular at $\psi = \pi/2$. 
For $C \neq 0$ we can solve $\p_{\psi}W_{\theta} = f(\eta)g(\psi)$ for $W_{\theta}$,
in order to obtain
\be
W_{\theta} \= -\frac{AB}{C}(\cosh\eta \, \cos\psi)^{C}   + \text{const} \,.
\ee
Now, the boundary condition at~$\eta \to \infty$ implies that~$C < 0$, but this leads to singularities at~$\psi = \pi/2$. 
Thus we find that there is no smooth solution for the~$p=0$ mode of~$W_\theta$.

Now we turn to the~$p \neq 0$ modes. In this case we could not easily solve the equation~\eqref{Wpqeqn} before 
imposing boundary conditions and smoothness. However, we can still show that there are no smooth solutions. 
To see this we expand the field~$W^{(p,q)}_{\theta}$ in a power series expansion near~$\eta=0$.
Imposing smoothness at~$\eta=0$\,\footnote{Near $\eta=0$, the AdS$_{2}$ part of the metric reduces to 
$ds^{2}_{\text{AdS}_{2}}=d\eta^{2}+\eta^{2}d\theta^{2}=dz\,d\bar z$ where $z=\eta e^{i\theta}$. By smoothness 
at $\eta=0$, we mean that the fields should be real analytic functions of $z$ and $\bar z$ so that all the derivatives of 
fields exist at $z=\bar z=0$. For example, smoothness of a scalar field~$\phi$ at the origin implies that its leading behavior  
is~$\phi(z,\bar z)=z^{a}\bar z^{b}$ for some non-negative integers~$a,b$. 
In the~$\eta,\theta$ coordinates this means that~$\phi(\eta) = \eta^{a+b} e^{i(a-b) \theta} = \eta^{|p|+r} e^{i p \theta} $,
where~$p=a-b$ and~$r=2\,\text{min}(a,b)\ge 0$.
Similarly the smoothness at $\eta=0$ of the gauge field (which are differential 1-forms)  
requires that $W_{\eta}\sim \eta^{|p|-1}$ and $W_{\theta}\sim \eta^{|p|}$ with~$p\neq 0$. This condition has been
used in a similar context in~\cite{David:2016onq}.}, 
\be \label{Eqn:expWtheta}
W^{(p,q)}_{\theta} (\eta, \psi) \= \eta^{|p|} \, \sum_{n\ge 0} \eta^{n} \, a_{n}(\psi) \,.
\ee
Using this expansion in Equation~\eqref{Wpqeqn}, we obtain coupled differential equations for the modes~$a_{n}(\psi,\phi)$. 
The equation for~$a_0$ is (we use the notation~$' \equiv \p_\psi$ below),
\be
\frac{a_0'}{a_0}  \=  \frac{q}{\sin \psi}  \,,
\ee
which can be solved easily to obtain
\be
a_0 \= C_0 \, (\tan(\psi/2))^q \,.
\ee
Since this function is singular in the domain~$\psi \in [0,\pi]$, we conclude that the only smooth solution has~$C_0=0$. 
With this solution we find that the next coefficient satisfies
\be
\frac{a_1'}{a_1} \=  \frac{pq}{|p|+1} \frac{1}{\sin \psi}  \,
\ee
with the solution, as above,
\be
a_1 \= C_1 \, (\tan(\psi/2))^{pq/(|p|+1)}\,,
\ee
which, once again, implies that the only smooth solution is~$a_1=0$.
Continuing in this manner we show in~Appendix \ref{App:nth}  that if the coefficients~$a_0=0,\dots,a_{\ell-1} =0$, the~$\ell^\text{th}$ 
term satisfies the equation
\be\label{Eqn:nthterm}
\frac{a'_\ell}{a_\ell} \=  \frac{pq}{|p|+\ell} \frac{1}{\sin \psi}\, .
\ee
As above, the only solution is~$a_\ell = C_\ell \, (\tan(\psi/2))^{pq/(|p|+\ell)}$, from which we conclude that~$a_\ell$ 
should also vanish if it is smooth.

Now that we have~$W_\theta=0$ as the only smooth solution, we can show that the other fields also vanish. 
In order to see that we substitute~$W_\theta=0$ into the first and fourth equation of \eqref{Eqn:fcow5} to obtain 
\be
\p_{\psi}W_{\eta} \= 0\,, \quad k_{+} \sin\psi \= \p_{\rho}W_{\eta}\,.
\ee
From this equation we see that~$k_{+} = 0$ is the only smooth solution, and that $W_{\eta}$ is independent of all the 
coordinates except~$\eta$. This implies that that 
\be
f_{\mu\nu} \= 0\,, \quad k_{+} \= 0\,,
\ee
and we can do a gauge transformation only depending on~$\eta$ in order to set~$W_{\eta}=0$.

\subsubsection*{Reality Condition A2}
We now work under reality condition A2, $W_{\rho}$ is purely imaginary and the $5d$ gauge fields~$W_\mu$ are real. 
The remaining fields, $\Sigma, k_{\pm}, k_{3}$ are also taken to be real.

The imaginary part of \eqref{Eqn:SPT2A2} is $\p_{\eta}W_{\rho} = 0$. As all fields must vanish as $\eta \rightarrow \infty$, we have
\be
W_{\rho} \= 0\,.
\ee
Substituting this into the imaginary part of \eqref{Eqn:SPT2B} fixes $\Sigma$ as in \eqref{Eqn:4dresultsigma}. 
The imaginary terms in the remainder of the complex equations, \eqref{Eqn:SPT2A1} and \eqref{Eqn:SPT2A3}, 
give $k_{-} = 0$ and $k_{3}$ is determined by \eqref{Eqn:4dresultk3}.

For the real parts of the field strengths, we get the same constraint equations as \eqref{Eqn:fcow5}. 
These equations do not have non-trivial smooth solutions as shown and therefore we get
\be
f_{\mu\nu} \= k_{+} \= 0\,.
\ee

\subsubsection*{Reality Condition B}
We now consider the analytic continuation where all the fields, $W_{M}, \Sigma, k_{\pm}, k_{3}$, are real.

The field strengths~$f_{\mu\nu}$ in the four-dimensional part of the metric can now be expressed 
in terms of~$k_+$ and~$f_{\mu \rho}$ as follows
\begin{equation}
\begin{split}
\ f_{\theta\eta} &\= \frac{\bigl(k_{+}\sin\psi +f_{\eta\rho}\bigr)\sinh^{2}\eta\sin 2\psi + f_{\psi\rho}\sinh 2\eta\sin^{2}\psi}{\bigl(\cos 2\psi -\cosh 2\eta\bigr)\sin\psi}\,, \\
\ f_{\theta\psi} &\= -\frac{\bigl(k_{+}\sin\psi + f_{\eta\rho}\bigr)\sinh 2\eta\sin^{2}\psi - f_{\psi\rho}\sinh^{2}\eta\sin 2\psi}{\bigl(\cos 2\psi -\cosh 2\eta\bigr)\sin\psi}\,, \\
\ f_{\theta\phi} &\= 0\,, \\
\ f_{\eta\psi} &\= -\frac{f_{\theta\rho}}{\tanh\eta\sin\psi} \,.
\label{Eqn:fcoreal}
\end{split}
\end{equation}
As all fields are real, the real part of the set of equations \eqref{EqnsSetB} give \eqref{Eqn:Kf}. For our Killing vector, the only non-zero components are $K^{\theta} = K^{\phi} = 1$ and so we obtain
\begin{equation} \label{fproj0}
\ f_{M\theta} \= f_{M\phi}\,.
\end{equation}
The equations~\eqref{Eqn:fcoreal}, \eqref{fproj0} can be summarised by
\begin{equation} \label{fconstraint}
f_{MN} - \frac{1}{2}\varepsilon_{MNRST}f^{RS}K^{T} \= e^{-i\bigl(\theta + \phi\bigr)}k_{+}f^{-1}X^{1}_{MN}\, .
\end{equation}
These equations have non-trivial smooth solutions which were discussed in~\cite{Gupta:2015gga}. 

From the imaginary part of \eqref{Eqn:SPT2B} we can see that $\Sigma$ is given by \eqref{Eqn:4dresultsigma}. 
From the imaginary parts of \eqref{Eqn:SPT2A1} and \eqref{Eqn:SPT2A3} we get $k_{-} = 0$ and $\Sigma = 2k_{3}\cosh\eta$. 
Thus we obtain~\eqref{Eqn:4dresultk3} as the solution for~$k_3$ once again.

\subsubsection*{Reality Condition C}
We now work under the analytical continuation where $f_{\mu\rho}$ is purely imaginary 
and all other fields,~$f_{\mu\nu}, \Sigma, k_{\pm}, k_{3}$, are real. 
The real terms of the complex equations, \eqref{Eqn:SPT2A1}--\eqref{Eqn:SPT2B}, 
can be rearranged to yield the following equations for the real parts of the field strengths
\begin{equation}\label{Eqn:fvi}
\begin{split}
\ f_{\theta\eta} & \= -f_{\eta\phi} \=\frac{k_{+}\sinh^{2}\eta\sin 2\psi}{\cos 2\psi -\cosh 2\eta}\,,  \\
\ f_{\theta\psi} & \= -f_{\psi\phi} \= -\frac{k_{+}\sinh 2\eta\sin^{2}\psi}{\cos 2\psi -\cosh 2\eta}\,,  \\
\ f_{\theta\phi} & \= 0\,, \\
\ f_{\eta\psi} & \= 0\,. \\
\end{split}
\end{equation}
This set of equations is precisely the equations of the 4d problem,
and we can use the same method as in~\cite{Gupta:2012cy} to evaluate~$f_{\theta \eta}$ and~$f_{\theta\psi}$. 
We define a new variable~$K$ through 
\begin{equation} \label{Eqn:Norho}
K \= \frac{k_{+}\sinh\eta\sin\psi}{\cos 2\psi -\cosh 2\eta} \,.
\end{equation}
The Bianchi identity in the $\theta\eta\psi$ directions gives
\begin{equation}
\bigl(\cot\psi\partial_{\psi} + \coth\eta\partial_{\eta}\bigr)K \= 0 \,.
\end{equation}
This implies that
\begin{equation}
K \= K\bigl(\cosh\eta\cos\psi, \theta, \phi, \rho\bigr) \,.
\end{equation}
As before we write this as a power series in the first variable we see that the boundary conditions imply that we only have 
negative powers of~$\cosh\eta\cos\psi$, which is then singular at the points where~$\cos \psi$ vanishes. 
Thus the boundary conditions and smoothness conditions imply that~$K = k_{+} = 0$. 
From~\eqref{Eqn:fvi} we see that~$f_{\theta \eta}=0$ and~$f_{\theta\psi}=0$ and thus all~$f_{\mu\nu} = 0$. 

The imaginary part of the basic equation \eqref{Eqn:SPT2A2} gives $f_{\eta\rho} = 0$. 
The Bianchi identies in the $\mu,\eta,\rho$ directions therefore reduce to 
\be
\p_{\eta} f_{\mu\rho} \= 0\,.
\ee
Applying the boundary conditions that~$f_{\mu\rho} \rightarrow 0$ as $\eta \rightarrow \infty$, we get $f_{\mu\rho} = 0$ everywhere. 
The remainder of the equations, \eqref{Eqn:SPT2A1}, \eqref{Eqn:SPT2A3} and \eqref{Eqn:SPT2B}, 
again yield the solutions \eqref{Eqn:4dresultsigma}, \eqref{Eqn:4dresultk3}, and $k_{-} = 0$.

\subsection{Vector Multiplet fluctuations around~$AdS_{2}\times S^{2} \ltimes S^{1}$  (spinning black hole)}
The bilinears corresponding to our Killing spinor~\eqref{KillSpinChosen} for the generic spinning black hole are summarized in Table~\ref{Tab:Spinbil}.
\begin{table}[h]
\centering
\begin{tabular}{ |c|c| } 
 \hline
 \textbf{ } & \textbf{Spinning black hole: Killing spinor Bilinears} \\
 \hline
 $f$ & $f = -2\bigl(\cosh\alpha\cosh\eta \+ \sinh\alpha\cos\psi\bigr)$  \\ 
 \hline
 $K^{A}$ & $K^{1} = 4\sinh\eta , \hspace{2mm} K^{4} = -4\sin\psi ,\hspace{2mm}K^{5} = -4\bigl(\sinh\alpha\cosh\eta \+ \cosh\alpha\cos\psi\bigr)$ \\ 
\hline 
  & $A^{21}_{12} = -2i\bigl(\cosh\alpha + \cos\psi\cosh\eta\sinh\alpha\bigr) , \quad A^{21}_{13} = 2i\sinh \alpha\sinh\eta\sin\psi$ , \\
$A^{12}_{AB}$ & $ A^{21}_{24} = 2i\cosh \alpha\sinh\eta\sin\psi ,$ \quad $A^{21}_{25} = 2i\cos\psi\sinh\eta ,$ \\
{} & $ A^{21}_{34} = -2i\bigl(\sinh\alpha + \cos\psi\cosh\eta\cosh\alpha\bigr) ,$ \quad $A^{21}_{35} = -2i\sin\psi\cosh\eta \,.$ \\
 \hline 
& $A^{11}_{12} =2e^{i(\theta+\phi)}\sinh\alpha\sinh\eta\sin\psi\,, \quad
A^{11}_{13} = 2e^{i(\theta+\phi)}\bigl(\sinh\alpha\cosh\eta\cos\psi +2 \cosh\alpha\bigr)\,,$ \\
 {} & $A^{11}_{14} = 2ie^{i(\theta+\phi)}\bigl(\sinh\alpha \cosh\eta+\cosh\alpha\cos\psi\bigr) \,, \quad A^{11}_{15} = -2ie^{i(\theta+\phi)}\sin\psi \,,$\\
$A^{11}_{AB}$  & $ A^{11}_{23} = -2ie^{i(\theta+\phi)}\bigl(\cosh\alpha\cosh\eta+\sinh\alpha \cos\psi\bigr) \,, \quad A^{11}_{24} = 2e^{i(\theta+\phi)}\bigl(\cosh\alpha\cosh\eta\cos \psi+\sinh\alpha\bigr) \,,$ \\ 
{} & $A^{11}_{25} = -2e^{i(\theta+\phi)}\sin\psi\cosh\eta, \quad
A^{11}_{34} = -2e^{i(\theta+\phi)}\sin\psi\sinh\eta\cosh\alpha\,,$ \\
{} & $A^{11}_{35} = -2e^{i(\theta+\phi)}\cos\psi\sinh\eta , \quad A^{11}_{45} = -2ie^{i(\theta+\phi)}\sinh\eta \,.$\\
\hline
\end{tabular}
\caption{Independent non-zero spinor bilinears for the Killing spinor $\epsilon^{1}_{++}$ of $AdS^{2}\times S_{2}\ltimes S_{1}$.}
\label{Tab:Spinbil}
\end{table}
\ndt Our first set of basic equations~\eqref{EqnsSetA}, in the spinning black hole context, are the following,
\begin{equation} \nn
\begin{split}
& \partial_{1}\Sigma\sinh\eta -\partial_{4}\Sigma\sin\psi -\partial_{5}\Sigma\bigl(\sinh\alpha\cosh\eta +\cosh\alpha\cos\psi\bigr) = 0 \,, 
\\ 
& i\Sigma\cosh^{2}\alpha + i\sinh\alpha\bigl(\Sigma\sinh\alpha +if_{34}-2k_{3}\cos\psi\bigr) \\
& \qquad \= \sinh\eta\bigl(f_{13}\sinh\alpha\sin\psi + f_{25}\cos\psi\bigr) -\cosh\eta \bigl(f_{12}\sinh\alpha\cos\psi + f_{35}\sin\psi\bigr)  \\ 
& \qquad\qquad - \cosh\alpha\bigl(f_{12}-f_{24}\sinh\eta\sin\psi+\cosh\eta\bigl(2i\Sigma\sinh\alpha\cos\psi -f_{34}\cos\psi - 2ik_{3}\bigr)\bigr)\,, 
\end{split}
\end{equation}
\begin{equation}\label{Eqn:SpA0}
\begin{split}
& k_{+}\bigl(\cosh\alpha\cosh\eta + \sinh\alpha\cos\psi\bigr)-i\Sigma\sinh 2\alpha\sinh\eta\sin\psi \\
&\qquad \=f_{24}\sinh\alpha -f_{35}\sinh\eta\cos\psi +\cosh\alpha\bigl(f_{13}+f_{24}\cosh\eta\cos\psi -f_{34}\sinh\eta\sin\psi\bigr) \\ 
&\qquad\qquad +f_{12}\sinh\alpha\sinh\eta\sin\psi +\cosh\eta\bigl(f_{13}\sinh\alpha\cos\psi-f_{25}\sin\psi\bigr) \,, \\ 
& ik_{-}\bigl(\cosh\alpha\cosh\eta + \sinh\alpha\cos\psi\bigr)\\
&\qquad  \= f_{15}\sin\psi +f_{45}\sinh\eta + \cosh\alpha\bigl(f_{23}\cosh\eta - f_{14}\cos\psi\bigr)+\sinh \alpha\bigl(f_{23}\cos\psi - f_{14}\cosh\eta\bigr)\,.
\end{split}
\end{equation}
Our second set of basic equations~\eqref{EqnsSetB}, in the spinning black hole context can be summarized as follows, 
\be\label{Eqn:SpB1}
-i\p_{M}\bigl(\Sigma\cosh\eta\cosh\alpha + \Sigma\cos\psi\sinh\alpha\bigr) \= f_{\theta M} + f_{M\phi} + \tanh\alpha f_{M\rho}\,.
\ee

Now we impose the various reality conditions. 
The analysis is similar to the static case, and therefore we will not present all the details and focus on the new points. 
The results are summarised in Table \ref{Tab:Spinsol}.

\begin{table}[h]
\centering
\begin{tabular}{ |c|c|c|c|c| } 
 \hline
& \textbf{A1}  & \textbf{A2}  & \textbf{B}  & \textbf{C} \\[-2mm]
 & {$W_{\rho}$ Im, $A_\mu$ Real} & {$W_{\rho}$ Im}, $W_\mu$ Real & {$W_{\rho}$ Real}  & {$f_{\mu\rho}$ Im} \\
 \hline
 \hline
 $\Sigma$ &$C  \cosh\alpha \, \text{sech}\eta$ & $0$ & $0$ & $0$ \\
 \hline
 $W_{\rho}$ & $-i C  \cosh\alpha \, \sinh\alpha \, \text{sech}\eta$   & $0$ & Solutions to \eqref{Eqn:fconstraint}  & $0$  \\
 \hline
 $\widetilde{f}_{\mu\nu}$, $k_+$ & 0, 0 & 0, 0 & Solutions to \eqref{Eqn:fconstraint}  & 0, 0\\ 
 \hline  
 $k_{-}$ & 0 & 0 & 0 & 0 \\ 
 \hline 
 $k_{3}$ & $(C/2) \, \text{sech}^2 \eta$  & $0$ & $0$ & $0$ \\ 
\hline
\end{tabular}
\caption{The complete set of off-shell BPS solutions for vector multiplet fluctuations around the near-horizon 
spinning BH for different reality conditions~A and~B. As before, 
$f_{MN}$ is the fluctuation of the field strength of~$W_M$, 
$\wt f_{\m\n}$ is the fluctuation of the field strength of~$A_{\m}$, $\Sigma$ is the 
fluctuation of the scalar~$\sigma$, and~$k_\pm, k^3$ are the fluctuations of the auxiliary fields~$Y^{ij}$. 
For condition~B, some explicit solutions to Equations \eqref{Eqn:fconstraint} are presented in~\cite{Gupta:2015gga}.}
\label{Tab:Spinsol}
\end{table}
Note that for reality condition A1, we can summarise the solutions as
\be
\begin{split}
\Sigma\cosh\alpha - iW_{\rho}\tanh\alpha & \= \frac{C}{\cosh\eta}\,, \\
\Sigma\sinh\alpha - iW_{\rho} & \= 0\,,
\end{split}
\ee
which is simply a rotation in field space of~$\Sigma$ and~$W_\rho$ of the static black hole solution.

\subsubsection*{Reality Condition A1} 
Her we have that $W_{\rho} = W_{5}\cosh\alpha$ is purely imaginary, 
the $4d$ gauge fields~$A_\mu$ are real, and the remaining fields are real.  
We start by decomposing the gauge fields in $5d$ in terms of the $4d$ gauge fields $A$ as done in 
\eqref{Eqn:AdS2S2S1euc} for the spinning black hole. In particular we have
\be\nn
B \= \cos\psi d\phi - \tanh\alpha (\cosh\eta - 1) d\theta \,,
\ee
\be
W_{\theta} \= A_{\theta} -\tanh\alpha\bigl(\cosh\eta -1\bigr)W_{\rho}  \,, \quad  
W_{\eta} \= A_{\eta} \,,\quad
W_{\psi} \= A_{\psi} \,,    \quad
W_{\phi} \= A_{\phi} + \cos\psi \, W_{\rho}\,.  
\ee
The imaginary part of~$M=2$, $M=3$ equations  
in \eqref{Eqn:SpB1} now give
\be \label{SigmaW5}
(\Sigma \cosh\alpha - iW_{\rho}\tanh\alpha)\cosh\eta + (\Sigma \sinh\alpha - W_{\rho})\cos\psi \= C(\theta, \phi, \rho)\,,
\ee
From the imaginary part of the~$M=1$ equation in~\eqref{Eqn:SpA0} we obtain
\be
2\Sigma\sinh\alpha\cosh^2\alpha \= -i(\coth\eta\, \p_{\eta} + \cot\psi \, \p_{\psi} - \cosh 2\alpha)W_{\rho}\,.
\ee
Combining these equations, we get
\be \label{Eqn:HJdif}
(\coth\eta\, \p_{\eta} + \cot\psi\, \p_{\psi} - 1)J\cosh\alpha \= (\coth\eta\p_{\eta} + \cot\psi\, \p_{\psi} + 1)H\sinh\alpha\,,
\ee
where
\be
H \= \Sigma\cosh\alpha - iW_{\rho}\tanh\alpha\,, \quad \quad J \= \Sigma\sinh\alpha - iW_{\rho}\,.
\ee
From \eqref{SigmaW5}, we also have
\be \label{Eqn:HJ}
H \cosh\eta + J\cos\psi \= C.
\ee
In order to solve these equations, we assume that the fields $H, J$, and the constant $C$ are smooth and admit a Taylor expansion
around~$\a=0$ for any value of the spacetime coordinates, i.e.,  
\be
H \= \sum_{n=0}^\infty H_{n} \, \a^n \,, \qquad  
J \= \sum_{n=0}^\infty J_{n} \, \a^n\,, \qquad
C \= \sum_{n=0}^\infty C_{n} \, \a^n\,.
\ee
At zeroth order in $\alpha$ this reduces to the non-spinning case for which we conclude
\be
H_{0} \= \frac{C_{0}}{\cosh\eta} \,, \qquad J_0 \=0\,.
\ee
At first order in $\alpha$ Equation~\eqref{Eqn:HJdif} leads to
\be
(\coth\eta\, \p_{\eta} + \cot\psi\, \p_{\psi} - 1)J_{1}  \= 0\,,
\ee
which is the same equation as that obeyed by the non-spinning variable~$J_0$. 
Thus we can use the same analysis to obtain~$J_1=0$.
Substituting this into \eqref{Eqn:HJ} we get the following equation for the first-order terms,
\be
H_{1} \= \frac{C_{1}}{\cosh\eta} \,.
\ee
Iteratively we can see that at the $n^\text{th}$ order,
\be
J_{n} \= 0\,, \quad
H_{n} \= \frac{C_{n}}{\cosh\eta}\,.
\ee
Thus we conclude that $J = 0$ and so $H = C \, \text{sech}\eta$. This leads to
\be
\Sigma \= \frac{C\cosh\alpha}{\cosh\eta}\,, \quad \quad W_{\rho} \= -i\frac{C\sinh\alpha\cosh\alpha}{\cosh\eta}\,.
\ee
Substituting these into the remaining equations we obtain $C$ to be a constant and 
\be
k_{-} \= 0\,, \quad\quad k_{3} \= \frac{C}{2\cosh^{2}\eta}\,. 
\ee

The real parts of Equations~\eqref{Eqn:SpB1} lead to
\be 
\widetilde{f}_{\mu\phi} \= -\widetilde{f}_{\theta\mu} + \p_{\rho}A_{\mu}\tanh\alpha\,, 
\qquad \p_{\rho}A_{\phi} \= \p_{\rho}A_{\theta}\,,
\ee
where $\widetilde{f}_{\mu\nu} = \p_{\mu} A_{\nu} - \p_{\nu} A_{\mu}$. Substituting these into the real part of \eqref{Eqn:SpA0} lead to the following three equations,
\be \label{fcowW5alpha1}
\begin{split}
& \bigl(\widetilde{f}_{\theta\eta}\coth\eta + \widetilde{f}_{\theta\psi}\cot\psi - \frac{\p_{\rho}A_{\psi}}{\sin\psi}\bigr)\coth\eta \\
& \quad + \tanh\alpha\left(\frac{\cos\psi}{\sinh\eta}\bigl(\widetilde{f}_{\theta\eta}\coth\eta + \widetilde{f}_{\theta\psi}\cot\psi - \frac{\p_{\rho}A_{\psi}}{\sin\psi}\bigr) - \cosh\eta\bigl(\frac{\p_{\rho}A_{\eta}}{\cosh\eta + 1} + \frac{\p_{\rho}A_{\psi}}{\sinh\eta\tan\psi}\bigr)\right) \\
& \quad \quad - \tanh^{2}\alpha\cos\psi\bigl(\frac{\p_{\rho}A_{\eta}}{\cosh\eta + 1} + \frac{\p_{\rho}A_{\psi}}{\sinh\eta\tan\psi}\bigr) \= 0\,,
\end{split}
\ee
\be\label{fcowW5alpha2}
\begin{split}
&\bigl(k_{+} + \widetilde{f}_{\theta\eta}\cot\psi - \widetilde{f}_{\theta\psi}\coth\eta - \frac{\p_{\rho}A_{\eta}}{\sin\psi}\bigr)\cosh\eta \\
& \quad + \tanh\alpha\left(\cos\psi\bigl(k_{+} + \widetilde{f}_{\theta\eta}\cot\psi - \widetilde{f}_{\theta\psi}\coth\eta - \frac{\p_{\rho}A_{\eta}}{\sin\psi}\bigr) - \cosh\eta\bigl(\frac{\p_{\rho}A_{\eta}}{\tan\psi} + \frac{(1-\cosh\eta)\p_{\rho}A_{\psi}}{\sinh\eta}\bigr)\right) \\
& \qquad - \tanh^{2}\alpha\cos\psi\bigl(\frac{\p_{\rho}A_{\eta}}{\tan\psi} - \frac{\sinh\eta\, \p_{\rho}A_{\psi}}{\cosh\eta + 1}\bigr) \= 0\,,
\end{split}
\ee
and
\be\label{fcowW5alpha3}
\widetilde{f}_{\eta\psi} \= \frac{\p_{\rho}A_{\theta}}{\sin\psi\tanh\eta} + \tanh\alpha\left(\frac{\p_{\rho}A_{\theta}}{\tan\psi\sinh\eta} \right)\,.
\ee
We can solve these three equations perturbatively in~$\alpha$ as we did above for the fields~$H$ and~$J$, by 
expanding the fields~$A_\mu$ and~$k_{+}$ in a series around~$\alpha=0$. 
Exactly as in the analysis below~\eqref{Eqn:HJ}, 
at each order~$n$ in this expansion, the fields satisfy the equations~\eqref{fcowW5alpha1}--\eqref{fcowW5alpha3} 
with~$\alpha=0$. 
These equations are simply the equations for the non-spinning case~\eqref{Eqn:fcow5}, 
for which we have already seen that there are no non-trivial smooth solutions. 
The final conclusion is that $k_{+} = A_{\mu} = 0$ are the only solutions.

\subsubsection*{Reality Conditions A2, B and C}

For the reality condition A2, $W_{\rho}$ is purely imaginary, the $5d$ gauge fields~$W_\mu$ are real, and 
the remaining fields $\Sigma, k_{\pm}, k_{3}$ are  real.
Using a similar analysis as above we find that~$W_{\rho} = \Sigma = k_{-} = k_{3} = 0$.
The remaining equations and reality conditions are the same as in the case~$A1$, for which
we already concluded that there are no non-trivial solutions.
Therefore all fields vanish for the choice A2.

For Condition B, i.e.~all fields real, the imaginary part of the third equation in \eqref{Eqn:SpA0} yields
\be
\Sigma \sinh 2\alpha\sin\psi\sinh\eta \= 0\,.
\ee
One solution is when $\alpha = 0$ for which we get back the static case.
The other set of solutions is $\Sigma = 0$. By the imaginary part of the second equation and the fourth equations 
in \eqref{Eqn:SpA0}, we get $k_{3} = k_{-} = 0$. The remainder of the equations are equivalent to the 
constraints~\eqref{Eqn:fconstraint} as in the static case. 
We conclude that $\Sigma = k_{3} = k_{-} = 0$. We also find that $f_{MN}$ and~$k_{+}$ obey 
the contact-instanton like equations~\eqref{Eqn:fconstraint}. These equations have non-trivial smooth 
solutions which were discussed in~\cite{Gupta:2015gga}.

Condition C is that~$f_{\mu\rho}$ imaginary and all other fields,~$f_{\mu\nu}, \Sigma, k_{\pm}, k_{3}$, are real. 
We can solve the real parts of the equations to obtain
the same equations as those obtained for the analytic continuation $f_{\mu\rho}$ imaginary in the static case, 
i.e.~Equations~\eqref{Eqn:fvi}. 
By the same analysis of the real terms in the Bianchi Identity in the $4d$ directions, we obtain $f_{\mu\nu} = k_{+} = 0$.
Solving the imaginary parts of the equations, we obtain $f_{25} = f_{35} = 0$. 
The second equation in \eqref{Eqn:SpA0} now gives $\Sigma = 0$. 
Substituting these values into the other equations leads to $k_{+} = k_{3} = f_{15} = f_{45} = 0$.
Thus all field fluctuations vanish in this case.

\section{Discussion \label{sec:Discussion}}

Our focus in this paper was spinning black holes in five-dimensional asymptotically flat space, defined 
as solutions to five-dimensional~$\CN=2$ supergravity. 
Upon reduction to four dimensions, one gets 
a supersymmetric non-spinning black hole in four-dimensional~$\CN=2$ supergravity with non-zero
electromagnetic flux in the Kaluza-Klein vector multiplet.  
The~4d/5d lift shows that the 4d black hole is the same as the 5d black hole at the center of Taub-NUT space. 
Thus the near-horizon limit of the 4d and 5d black holes are the same. 
Since the quantum entropy is defined purely 
in the near-horizon configuration, the expectation is that the quantum entropies of the 
4d and the 5d black holes are equal.

Our main results in this paper concern the localization manifold for the functional integral for the quantum entropy of the 5d black hole.
These results are consistent with the above expectation. The set of off-shell 5d vector multiplets BPS solutions is precisely the lift of the 
corresponding 4d set, and we show that there are no new solutions in the Kaluza-Klein sector, i.e.~solutions that depend 
on the compactification direction. This is a bit of a surprise from the point of view of the actual analysis as we have the same 
number of equations, and all the fields a priori could have a non-trivial dependence on the fifth direction. 
In addition to this main result we also have some results on the localization manifold in the Weyl multiplet sector and 
also the result of different choices of analytic continuation. We publish these technical results as they may be 
useful in related computations.

The next step in the quantum entropy program for 5d spinning black holes is clear, we need to calculate the action and 
one-loop determinant and assemble all the pieces in the localization formula. This should give the quantum entropy of 
the 5d BH at all orders in perturbation theory---a 5d analog of the OSV formula---perhaps 
along the lines of~\cite{Dijkgraaf:2004te}. These investigations are being pursued, and we hope to report on this soon.

%


\section*{Acknowledgements}

We thank Chris Couzens and Atish Dabholkar for useful conversations. 
This work was supported by the ERC Consolidator Grant N.~681908, ``Quantum black holes: A macroscopic window into the 
microstructure of gravity'', and by the STFC grant ST/P000258/1.

\appendix
\section{Conventions and useful identities for spinor algebra}\label{Conv.SpinorAlg.}

Our five-dimensional Euclidean gamma matrices are as follows.
\begin{equation}
\begin{split}
\gamma_{1} \=\begin{pmatrix}
 1 & 0 & 0 & 0 \\
 0 & 1 & 0 & 0 \\
 0 & 0 & -1 & 0 \\
 0 & 0 & 0 & -1 \\
\end{pmatrix}
\,, \qquad
\gamma_{2} \=\begin{pmatrix}
 0 & 0 & 1 & 0 \\
 0 & 0 & 0 & 1 \\
 1 & 0 & 0 & 0 \\
 0 & 1 & 0 & 0 \\
\end{pmatrix}\,, \\
\gamma_{3}\=\begin{pmatrix}
 0 & 0 & 0 & i \\
 0 & 0 & i & 0 \\
 0 & -i & 0 & 0 \\
 -i & 0 & 0 & 0 \\
\end{pmatrix}
\,, \qquad
\gamma_{4}\=\begin{pmatrix}
 0 & 0 & 0 & 1 \\
 0 & 0 & -1 & 0 \\
 0 & -1 & 0 & 0 \\
 1 & 0 & 0 & 0 \\
\end{pmatrix}
\,, \\
\gamma_{5}\=\begin{pmatrix}
 0 & 0 & i & 0 \\
 0 & 0 & 0 & -i \\
 -i & 0 & 0 & 0 \\
 0 & i & 0 & 0 \\
\end{pmatrix}\,.
\end{split}
\end{equation}

These matrices are Hermitian and satisfy the following properties
\begin{equation} \label{Eqn:gamma12com}
\begin{split}
\gamma_{MNPQR} & \= \mathbf{1}\varepsilon_{MNPQR} \,.
\\
\left[\gamma_{a},\gamma_{bc}\right] & \= 4\delta_{a[b}\gamma_{c]}\,.
\end{split}
\end{equation}
The charge conjugation matrix is 
\be
C \= 
\begin{pmatrix}
 0 & -1 & 0 & 0 \\
 1 & 0 & 0 & 0 \\
 0 & 0 & 0 & -1 \\
 0 & 0 & 1 & 0\\
\end{pmatrix} \,,
\ee
and obeys the following properties,
\begin{equation}
C^{T} \= - C\,, \hspace{10mm} C^{\dagger} \= C^{-1} \,, \hspace{10mm} \gamma_{\mu}^{T} \= C\gamma_{\mu} C^{-1} \,. 
\end{equation}
We define symplectic Majorana spinors, $\epsilon_{\alpha}^{i}$, $i = 1,2$, by 
\begin{equation}
\epsilon_{i} \= \varepsilon_{ij}\epsilon^{j} \,,
\end{equation}
where $\varepsilon_{ij}$ is antisymmetric with $\varepsilon_{12}=1$.
The conjugate of a spinor is defined by $\overline{\epsilon}_{i} = \epsilon^{i}{}^{\dagger}$, so that
\begin{equation}
\begin{split}
\ \overline{\epsilon}_{1} &\= \overline{\epsilon}^{2} = \epsilon^{1}{}^{\dagger}\,, \\
\ \overline{\epsilon}_{2} &\= -\overline{\epsilon}^{1} = \epsilon^{2}{}^{\dagger}\,.
\end{split}
\end{equation}
The symplectic Majorana condition is
\begin{equation}
C^{-1}\overline{\epsilon}_{i}^{T} = \varepsilon_{ij}\epsilon^{j}.
\end{equation}
Explicitly, the spinors and their conjugates are related by
\begin{equation}
\overline{\epsilon}^{1*} = \overline{\epsilon}^{2}C^{-1}, \hspace{5mm} \overline{\epsilon}^{2*} = - \overline{\epsilon}^{1}C^{-1}, \hspace{5mm} \epsilon^{1*} = C\epsilon^{2}, \hspace{5mm} \epsilon^{2*} = -C\epsilon{^1}.
\end{equation}

\section{Killing Spinors}\label{App:KS}
The Killing spinor equation can be rewritten as 
\begin{equation}
\begin{split}
\nabla_{M}\epsilon & \= -\frac{i}{4}T_{PN}\bigl(3\gamma^{PN}\gamma_{M}-\gamma_{M}\gamma^{PN}\bigr)\epsilon \,, \\
 &\= 2iT_{MN}\gamma^{N}\epsilon - \frac{i}{2}T^{PN}\gamma_{MPN}\epsilon.
\label{Eqn:KSEsimple}
\end{split}
\end{equation}

In the spinning case, the near-horizon configuration is given by the metric
\begin{equation}
ds^{2} \= \sinh^{2} \eta d\theta^{2} + d\eta^{2} + d\psi^{2} + \sin^{2}\psi d\phi^{2} + \cosh^{2}\alpha\bigl(d\rho + \cos\psi d\phi - \tanh\alpha(\cosh\eta - 1)d\theta\bigr)^{2}\,,
\end{equation}
and the only non-zero components of auxiliary field $T$
\begin{equation}
T_{\theta\eta} \= -\frac{i}{4}\sinh\eta\cosh\alpha\,, 
\quad T_{\psi\phi} \= \frac{i}{4}\sin\psi\sinh\alpha\,.
\end{equation}
Solving the Killing Spinor Equation \eqref{Eqn:KSEsimple}, with these values of $g_{MN}$ and $T$, we get that the Killing spinors must be 
\begin{equation}
\epsilon =  \exp\bigl(\frac{\alpha}{2}\bigr)\exp\bigl(-\frac{\eta}{2}\gamma_{1}\bigr)\exp\bigl(\frac{\theta}{2}\gamma_{21}\bigr)\exp\bigl(\frac{\psi}{2}\gamma_{45}\bigr)\exp\bigl(\frac{\phi}{2}\gamma_{34}\bigr)\epsilon_{0}\,,
\label{Eqn:KSAdS2S3T2}
\end{equation}
where $\epsilon_{0}$ is a constant spinor.
We label the components of the constant spinor as below,
\begin{equation}
\epsilon^{0} \equiv \bigl( a_{1}\,,
a_{2}\,,
a_{3}\,,
a_{4} \bigr)^{T}\,. \nn
\end{equation}

Choosing the constants to be 
\begin{equation}
\begin{split} \nn
\ \bigl(a_{1},a_{2},a_{3},a_{4}\bigr) & \= \bigl(1,0,I,0\bigr), \hspace{5mm} \bigl(a_{1},a_{2},a_{3},a_{4}\bigr) = \bigl(0,1,0,I\bigr), \\
\ \bigl(a_{1},a_{2},a_{3},a_{4}\bigr) & \= \bigl(0,1,0,-I\bigr), \hspace{5mm} 
\bigl(a_{1},a_{2},a_{3},a_{4}\bigr) = \bigl(1,0,-I,0\bigr)\,,
\end{split}
\end{equation}
respectively, we find four linearly independent Killing spinors
\begin{equation}
\begin{split}
\epsilon^{1}_{-+} & \= e^{-\frac{i}{2}\bigl(\theta -\phi\bigr)}  \begin{pmatrix}{c}
\cos\frac{\psi}{2}\big(\cosh\bigl(\frac{\alpha -\eta }{2}\bigr)-\sinh \bigl(\frac{\alpha +\eta }{2}\bigr)\bigr) \\
i\sin\frac{\psi}{2}\big(\cosh\bigl(\frac{\alpha +\eta }{2}\bigr)+\sinh \bigl(\frac{\alpha -\eta }{2}\bigr)\bigr) \\
i\cos\frac{\psi}{2}\big(\cosh\bigl(\frac{\alpha +\eta }{2}\bigr)-\sinh \bigl(\frac{\alpha -\eta }{2}\bigr)\bigr) \\ 
-\sin\frac{\psi}{2}\big(\cosh\bigl(\frac{\alpha -\eta }{2}\bigr)+\sinh \bigl(\frac{\alpha +\eta }{2}\bigr)\bigr)
\end{pmatrix} \,, 
\\
\epsilon^{1}_{+-} & \= e^{+\frac{i}{2} \bigl(\theta -\phi\bigr)} \begin{pmatrix}{c}
i\sin\frac{\psi}{2}\big(\cosh\bigl(\frac{\alpha -\eta }{2}\bigr)-\sinh \bigl(\frac{\alpha +\eta }{2}\bigr)\bigr) \\
\cos\frac{\psi}{2}\big(\cosh\bigl(\frac{\alpha +\eta }{2}\bigr)+\sinh \bigl(\frac{\alpha -\eta }{2}\bigr)\bigr) \\
-\sin\frac{\psi}{2}\big(\cosh\bigl(\frac{\alpha +\eta }{2}\bigr)-\sinh \bigl(\frac{\alpha -\eta }{2}\bigr)\bigr) \\ 
i\cos\frac{\psi}{2}\big(\cosh\bigl(\frac{\alpha -\eta }{2}\bigr)+\sinh \bigl(\frac{\alpha +\eta }{2}\bigr)\bigr)
\end{pmatrix}  \,, \\
\ \epsilon^{1}_{--} & \= e^{-\frac{i}{2} \bigl(\theta +\phi\bigr)}  \begin{pmatrix}{c}
i\sin\frac{\psi}{2}\big(\cosh\bigl(\frac{\alpha +\eta }{2}\bigr)+\sinh \bigl(\frac{\alpha -\eta }{2}\bigr)\bigr) \\
\cos\frac{\psi}{2}\big(\cosh\bigl(\frac{\alpha -\eta }{2}\bigr)-\sinh \bigl(\frac{\alpha +\eta }{2}\bigr)\bigr) \\
\sin\frac{\psi}{2}\big(\cosh\bigl(\frac{\alpha -\eta }{2}\bigr)+\sinh \bigl(\frac{\alpha +\eta }{2}\bigr)\bigr) \\ 
-i\cos\frac{\psi}{2}\big(\cosh\bigl(\frac{\alpha +\eta }{2}\bigr)-\sinh \bigl(\frac{\alpha -\eta }{2}\bigr)\bigr)
\end{pmatrix}  \,, 
\\
\epsilon^{1}_{++} &\= e^{+\frac{i}{2} \bigl(\theta +\phi\bigr)} \begin{pmatrix}{c}
\cos\frac{\psi}{2}\big(\cosh\bigl(\frac{\alpha +\eta }{2}\bigr)+\sinh \bigl(\frac{\alpha -\eta }{2}\bigr)\bigr) \\
i\sin\frac{\psi}{2}\big(\cosh\bigl(\frac{\alpha -\eta }{2}\bigr)-\sinh \bigl(\frac{\alpha +\eta }{2}\bigr)\bigr) \\
-i\cos\frac{\psi}{2}\big(\cosh\bigl(\frac{\alpha -\eta }{2}\bigr)+\sinh \bigl(\frac{\alpha +\eta }{2}\bigr)\bigr) \\ 
\sin\frac{\psi}{2}\big(\cosh\bigl(\frac{\alpha +\eta }{2}\bigr)-\sinh \bigl(\frac{\alpha -\eta }{2}\bigr)\bigr)
\end{pmatrix} \,.
\label{Eqn:KillingSpinorsT2}
\end{split}
\end{equation}

\section{Proof of Equation \eqref{Eqn:nthterm}}\label{App:nth}
We would like to prove that when we expand~$W_{\theta}$ as in~\eqref{Eqn:expWtheta} then 
the~$\ell^\text{th}$ term satisfies the equation~\eqref{Eqn:nthterm} if~$a_0=0,\dots,a_{\ell-1} =0$.

We begin by substituting \eqref{Eqn:expWtheta} into \eqref{Wpqeqn} and multiply across by $\cosh^2 \eta$ to obtain
\be
\begin{split} \label{Eqn:Wthetapn}
&\sum^{\infty}_{n = 0}\biggl(\bigl( -pq\sin\psi \, a_{n}\bigr) \eta^{|p| + n} + \frac{1}{2}\bigl((|p| + n)\sin^2\psi \, a'_{n}\bigr) \eta^{|p| + n -1}\sinh 2\eta \\
&\qquad \qquad \qquad \qquad+ \bigl(\sin\psi\, \cos\psi \, a''_{n} -pq\sin\psi \, a_{n} - a'_{n}\bigr) \eta^{|p| + n}\sinh^2 \eta \biggr) \= 0\,.
\end{split}
\ee
Using the following series expansions,
\be
\sinh^2 x \= \sum^{\infty}_{k = 1}\frac{2^{2k-1}}{(2k)!}\,x^{2k}\,, \qquad
\sinh 2x \= \sum^{\infty}_{k = 0}\frac{2^{2k+1}}{(2k+1)!}\,x^{2k+1}\,,
\ee
and multiplying across by~$\eta^{-|p|}$, we obtain
\be
\begin{split} \label{Eqn:Wthetanexp}
& \sum^{\infty}_{n = 0}\Bigl(\bigl( -pq\sin\psi \, a_{n}\bigr) \eta^{n} + \bigl((|p| + n)\sin^2\psi \, a'_{n}\bigr)\sum^{\infty}_{k = 0}\frac{2^{2k}}{(2k+1)!}\, \eta^{2k+n} \\
& \qquad \qquad \qquad+ \bigl(\sin\psi\, \cos\psi a''_{n} -pq\sin\psi \, a_{n} - a'_{n}\bigr) \sum^{\infty}_{k = 1}\frac{2^{2k-1}}{(2k)!} \, \eta^{2k + n} \Bigr) \= 0\,.
\end{split}
\ee
At order~$\ell$ in~$\eta$, we have
\be
\begin{split} \label{Eqn:Wthetal}
& -pq\sin\psi \, a_{\ell} + \sum^{\infty}_{k = 0}\frac{2^{2k}}{(2k+1)!}\bigl((|p| + \ell - 2k)\sin^2\psi\bigr) \, a'_{\ell - 2k}\\
& \qquad  \qquad + \sum^{\infty}_{k = 1}\frac{2^{2k-1}}{(2k)!}\bigl(\sin\psi\cos\psi\, a''_{\ell - 2k} -pq\sin\psi \, a_{\ell - 2k} - a'_{\ell - 2k}\bigr)\= 0\,.
\end{split}
\ee
Using the fact that all coefficients up to~$a_{\ell}$ vanish,
we see that the third term in~\eqref{Eqn:Wthetal} vanishes and the second term is zero except for~$k = 0$ and so~$a_{\ell}$ satisfies 
\be
\frac{a'_{\ell}}{a_{\ell}} \= \frac{pq}{(|p| + \ell)\sin\psi}\,,
\ee
which is precisely Equation~\eqref{Eqn:nthterm}.
%

\providecommand{\href}[2]{#2}\begingroup\raggedright\endgroup

\end{document}